# Developers' Experience with Generative AI Beyond Productivity Assessment – Insights from an Empirical Mixed-Methods Field Study

Developers' Experience with Generative AI Beyond Productivity

Charlotte Brandebusemeyer

Digital Health – Connected Healthcare, Hasso Plattner Institute, University of Potsdam, Potsdam, char.brandebusemeyer@hpi.de

Kerim Zunic

Hasso Plattner Institute, University of Potsdam, Potsdam, kerim.zunic@hpi.uni-potsdam.de

Thomas Zimmermann

University of California, Irvine, tzimmer@uci.edu

Tobias Schimmer

SAP Labs, SAP, Newport Beach, tobias.schimmer@sap.com

Bert Arnrich

Digital Health – Connected Healthcare, Hasso Plattner Institute, University of Potsdam, Potsdam, bert.arnrich@hpi.de

With the growing adoption of AI-powered coding assistants, organizations and developers are increasingly seeking to optimize their interaction with these tools. Prior research has largely focused on output quality and productivity gains, with limited attention paid to developers' well-being and interaction experiences. This paper presents a developer-centered empirical mixed-methods study to investigate how professional developers engage with Generative AI (GenAI) in their natural work environment. Controlled data collection sessions are combined with natural work periods. Results show that developers are generally satisfied with GenAI, particularly for monotonous, repetitive, and structured tasks, and report perceived efficiency and productivity gains. Copilot interaction type preferences differ by task type and complexity: While both in-code suggestions and chat-based prompting independently improve task efficiency and reduce perceived workload, combining these interaction types within a single task diminishes benefits. We propose a rule-of-thumb for selecting an interaction type based on task characteristics. During development-heavy tasks, results indicate that perceived cognitive load arises from AI interaction, while perceived productivity depends on AI output quality. Participation in this study positively influenced developers' awareness and intentional use of GenAI tools. These findings demonstrate the value of real-world, mixed-methods study designs to understand GenAI tools and developers' experiences with them.

## 1 INTRODUCTION

According to the Artificial Intelligence Index Report 2025 [35], 41.76% of all computer science publications worldwide published in 2023 were related to artificial intelligence (AI), indicating high academic interest in AI. However, this interest is not limited to academia. With increasing research indicating productivity gains through AI, industry invests substantially in AI [35]. Especially Generative AI (GenAI) has caught the interest of firms, as can be witnessed by increasing investments and usage [35]. However, according to Gartner's 2025 Hype Cycle for Artificial Intelligence [19], GenAI has entered the "trough of disillusionment", which is a phase characterized by the realization of initial overly high expectations regarding new technology. Systematic evaluations of people's interaction with GenAI, e.g., in a natural work setting, can reveal potentials and shortcomings of this new technology.

In the field of software engineering, AI-powered coding assistants like GitHub Copilot are utilized to automate programming-related tasks and to thereby relieve programmers and increase their productivity. This research builds on [7] findings, and their detected research gaps are going to be mentioned in the following: Previous research on GenAI has primarily focused on evaluating its output quality [26, 28, 42, 43] and developer's productivity gains [1, 9, 32, 45]. Although the technologies were developed for programmers, the human-centered perspective, including developers' behavior, interaction patterns and subjective experiences when working with GenAI, remains understudied. Recent studies have begun to address these aspects [1, 2, 9, 27, 32, 34, 38, 41, 45], yet empirical studies conducted in real-world organizational settings are still rare within academic literature [1, 9, 27, 32, 38]. Some of those studies rely solely on subjective data sources from questionnaires, interviews or blog posts [9, 34, 38] and thereby miss validation through objective data sources. Conducting mixed-methods studies with subjective and objective data sources and quantitative and qualitative data analyses can enhance generalizability and comprehension of the research results and can enable a nuanced analysis of the developer-GenAI interaction. Furthermore, besides [7], systematic research investigating how different interaction types (in-code suggestions vs chat prompts) and task types (coding, debugging, documentation, testing, summarizing, brainstorming) affect output quality, developers' interaction quality, and perceived workload in natural work settings are missing.

To address some of the above-mentioned research gaps, we conducted an empirical mixed-methods study involving professional software developers at SAP. Over a period of four days, we observed 22 developers in how they use GenAI tools in controlled and uncontrolled settings. GitHub Copilot was at the time of the study increasingly integrated into developers' workflows at SAP and was therefore chosen as GenAI tool during the controlled study setup. The developers were free to choose to interact with other GenAI tools during the uncontrolled study period. Our study integrates multimodal data sources, including subjective experience data (via questionnaires), behavioral data (from screen, mouse, and keyboard recordings), and physiological data (collected via a wristband) to gain fine-grained insights on the developer-GenAI interaction. By combining controlled and uncontrolled study phases, the study design balances experimental control (controlled tasks) with ecological validity by gathering realistic impressions of professional software developers' workdays within their daily work environment (uncontrolled work periods). In our previous work [7], we motivated and detailed the experimental design of our study and presented first findings from screen recording analyses (behavioral data) on the relevance of the way developers interact with GitHub Copilot during working tasks. This paper presents extended findings building on the results reported in [7]. Specifically, subjective questionnaire data gathered during the controlled sessions of the study are going to be analyzed in detail and brought into context with the behavioral data from the screen recordings. Furthermore, the results of the controlled sessions are brought into context with the subjective evaluations from the uncontrolled study phase. An overview of the empirical study and a summary of the research results from our previous work [7] are included in this paper to provide the necessary background information to interpret the results holistically.

With this study, we aim to address the following research questions:

**(RQ1)** How do software developers at SAP experience their interaction with AI/GenAI during their workday prior to the study?

**(RQ2)** Which GenAI interaction type is most frequently used and preferred by the participants during the controlled sessions of this study?

**(RQ3)** How does the interaction with GenAI during the controlled sessions of this study impact developers' efficiency, accuracy and perceived workload?

**(RQ4)** How is the interaction with AI evaluated during everyday working tasks concerning cognitive load and productivity?

**(RQ5)** What concerns and opportunities do developers foresee in the future with GenAI?

**(RQ6)** What was the impact of this study on developers' subsequent use of GenAI?

This empirical study contributes to emerging research that takes a developer-centered perspective to analyze the developer-GenAI interaction in a natural work environment within a firm. We see our contributions in:

- Proposing a feasible multimodal study design that combines controlled and uncontrolled study phases to evaluate developers' interaction with GenAI tools within a professional work environment
- Giving insights on how subjective and behavioral data combined give deeper insights into GenAI's impact on developers' experience and productivity
- Suggesting a rule-of-thumb for selecting an interaction type depending on the working task at hand

While RQ3 is addressed in depth in [7], parts of the results are also summarized in this paper. The remaining five research questions focus on developers' subjective experiences, providing contextual insights that complement the behavioral data analysis in RQ3 and put the findings into perspective. The following sections consist of the background and related work, an overview of the conducted empirical study, the analysis methodology of the results and the findings. The discussion section interprets the results in relation to the research questions and suggests a rule-of-thumb to select the appropriate interaction type for a task type. The validity of the conducted study and future research directions are discussed and the conclusion gives a final summary of this paper.

## 2 BACKGROUND

Software developers apply technical, problem-solving and communication skills on a daily basis and often whilst working on several projects simultaneously. The amount and diversity of deadline-oriented tasks can lead to a considerable mental burden, which can affect developers' mental health and lead to a decrease in productivity. With the growing availability of GenAI, both developers and organizations are exploring models and tools that can support developers during their work, enhance their productivity and streamline software development workflows.

Whilst organizations consider mainly productivity gains which lead to economic benefits, the developers' well-being when interacting with GenAI is also an important factor to consider. Studies have shown that well-being is associated with work satisfaction and productivity [13, 36]. Thus, the developer experience, i.e., how developers think about, feel and value their work [16] also contributes to productivity enhancement. The developer experience (DevEx) framework highlights three main factors that impact DevEx: cognitive load, flow state, and feedback loop [30]. Whilst we have focused on analyzing the cognitive load aspect in previous work [6], we examine the cognitive load aspect together with the feedback loop in this paper. In this context, we define the feedback loop as the interaction between developers and GenAI tools and examine the interaction in terms of the GenAI output quality, developer-GenAI interaction quality and developers' efficiency and general satisfaction during the interaction. The cognitive load, i.e. the mental processing required, in terms of perceived workload during tasks with or without the use of GenAI and different interaction types with GenAI are considered during the developer-GenAI interaction. By considering the feedback loop and cognitive load aspects of developer experience during developer-GenAI interaction, we gain insights surpassing mere productivity gains and can evaluate the benefits and disadvantages of the interaction in a developer-centered manner.

## 3 RELATED WORK

In the software engineering field, developments in AI programming assistants like GitHub Copilot are followed closely. As mentioned above, research has so far mainly focused on evaluating coding assistants regarding their output quality and productivity gains, whilst literature on developers' experiences during the interaction is scarce.

### 3.1 Output quality

Several empirical studies have systematically investigated the performance of AI code generation tools by assessing the correctness and quality of the generated code, mostly across standardised benchmarks. Yetistiren et al. [42] assessed the quality of generated code from Copilot on the HumanEval dataset, consisting of 164 code snippets in Python [8]. They found that in 91.5% of cases, Copilot generated valid code, but only 28.7% of the outputs were fully correct, with the rest being partially correct or incorrect. For correctly generated solutions, there was no significant difference in code efficiency between Copilot-generated code and human-generated canonical solutions. However, without human supervision, Copilot was less likely to generate correct code, whereas additional input from programmers increased the correctness of its output. In a following study, Yetistiren et al. [43] compared, among others, the success rate of the three prominent code generation tools GitHub Copilot, Amazon CodeWhisperer and OpenAI's ChatGPT, also using the HumanEval dataset. ChatGPT produced the most correct solutions (65.2% correct) out of the three. When using dummy function names to generate code ChatGPT also had the highest success rate with 61.6% of tasks solved correctly. Although Amazon CodeWhisperer and GitHub Copilot were not as successful, the authors found that both are rapidly improving, thereby acknowledging the currently fast progress in AI code generation tools.

Complementary insights were offered by Dakhel et al. [26], who focused on evaluating Copilot's output correctness and comparing it to students' solutions. While students produced more correct answers, Copilot's incorrectly generated code was easier to repair. The authors suggest that Copilot is beneficial for experienced programmers capable of critical evaluation of Copilot's output, whereas novices may trust the output and fail to detect suboptimal solutions.

Li et al. [21] compared Copilot's output across three different datasets (HumanEval, AixBench and MBPP). They found that while Copilot-generated code is functionally correct, it often performs worse than human-written code. The authors attribute performance regressions in Copilot-generated code to inefficient function calls, looping constructs, algorithms and use of language features, and highlight the value of precise prompt engineering to enhance performance.

Nguyen et al. [28] also examined the correctness of code generated by Copilot and considered how correctness and understandability vary across programming languages. They evaluated Copilot on 33 LeetCode tasks in four programming languages and found differences across languages, with Java achieving the highest correctness rate (57%).
All these studies share a common focus on evaluating GitHub Copilot's output correctness based on a set of predefined tasks. They include comparisons between tools, datasets, between tool and human output quality, and programming languages, thereby highlighting both the rapid improvement of models and tools and their remaining performance limitations. While these studies provide valuable and deep insights into GenAI output in controlled settings, they do not assess performance in natural work environments with real-world software engineering problems.

### 3.2 Productivity gains

Research investigating the impact of GenAI tools on developers' productivity is growing. The focus often lies on analyzing productivity gains when using GitHub Copilot. Studies differ in their use of mixed-methods designs, participant profiles and realism of study environments.

Bakal et al. [1] examined the effect of GitHub Copilot on professional developers' productivity within a firm. They combined quantitative metrics on Copilot use, including acceptance rates of suggestions and lines of code, with qualitative developer feedback collected. The average acceptance rate of Copilot's suggestions was 33%, but differences could be observed between programming languages with Go, Typescript, Python and Java having the highest acceptance rates. According to the subjective perception of most developers in the study, Copilot reduced the amount of time per task, enabled more tasks to be completed per sprint and improved work quality. Overall, developers were satisfied with Copilot, with a satisfaction rate of 72%.

Peng et al. [32] investigated GitHub Copilot's impact on productivity in terms of efficiency in a controlled experiment. Developers were asked to implement an HTTP server in JavaScript as quickly as possible. The group of developers that used Copilot was 55.8% faster than the control group that did not use Copilot. Survey responses showed that participants underestimated their actual productivity gains through Copilot, while the treatment group was willing to pay significantly more for a new Copilot release than the control group, indicating perceived benefits from Copilot use.

Coutinho et al. [9] particularly investigated software practitioners' perceived productivity when using AI tools in their pilot study. Participants reported positive aspects, such as quick access to information for learning and knowledge acquisition, time optimization, and the ability to work on a variety of software tasks with GenAI tools. Challenges included reliability issues with AI output, which require manual adjustments.

Ziegler et al. [45] examined if subjective perception of productivity correlates with the usage of GitHub Copilot. In a study with students and professionals, they found that the acceptance rate of Copilot suggestions showed the strongest correlation out of their chosen metrics with perceived productivity.

Overall, these studies consistently indicate that GenAI tools such as GitHub Copilot are associated with increased developer productivity, either in terms of measured efficiency gains or perceived productivity improvements. However, further research is needed to strengthen these findings, particularly through studies that combine subjective perceptions with objective usage or performance data to gain a more holistic understanding. In addition, further studies should be conducted in natural work environments within firms, involving professional developers and utilizing realistic software development tasks, to validate previous findings under real-world conditions. To date, there is no consensus on a standard productivity metric for evaluating interactions with GenAI tools - acceptance rates of code suggestions, efficiency and/or self-reported productivity are measures used. Moreover, research that explicitly examines productivity gains in relation to a positive developer experience and well-being when working with GenAI tools is scarce.

### 3.3 Developer-centered perspectives on developer-GenAI interaction

The studies presented above on productivity gains are taking a first step towards analyzing the interaction between developers and GenAI tools more closely. Beyond productivity, studies have begun to explore developer behavior, user experience, and satisfaction during GenAI-assisted programming.

The following five studies present insights into developers' behavior during the interaction with GenAI: Barke et al. [2] investigated developer behavior in a controlled experimental setting and identified two distinct interaction modes with GitHub Copilot: acceleration mode*,* where developers follow a clear plan and use Copilot to boost efficiency, and exploration mode*,* where they are uncertain how to proceed and explore options with Copilot. The authors suggest that Copilot's code suggestions should be tailored to these interaction modes: in acceleration mode and in a flow state, long suggestions with low confidence should not be shown, whilst in exploration mode, varied options can be helpful, but can also lead to cognitive overhead. Participants also highlighted a need for improved mechanisms to validate the quality of Copilot's outputs.

Mozannar et al. [27] also analyzed developers' interaction behaviour by identifying 12 activities that are related only to the interaction with a coding assistant like GitHub Copilot. They found that programmers spend a substantial portion of time reviewing and adapting Copilot suggestions and spend more than half their time on GenAI-related activities during tasks. Their results show a change in user behaviour when Copilot is integrated in an integrated development environment (IDE).

Stray et al. [38] analysed the developer-GenAI interaction in the broader context of a firm setting through interviews and observations. Participants appreciated the efficiency gains provided by GenAI but also reported challenges such as inaccurate responses and workflow disruptions. In terms of team dynamics, developers used GenAI to quickly find answers without disturbing colleagues, yet they still regarded their peers' responses as more reliable and valued their domain knowledge.

Tang et al. [40] adopted a multimodal research approach by combining IDE telemetry, eye-tracking data, and subjective workload assessments to investigate how developers validate and repair LLM-generated code. Their study revealed that awareness that the code was AI-generated significantly shaped developers' debugging behaviors and interaction strategies, influenced performance, and increased cognitive workload compared to cases where the code origin is not disclosed.

In our previous research [7], we also analyzed the developers' behavior during GitHub Copilot interaction and found that the interaction type (no Copilot, only in-code suggestions, only chat prompts or a combination of both) and intensity of its use to work on different simulated software engineering tasks had an impact on efficiency, accuracy and perceived workload of the developers.

The usability of GitHub Copilot was investigated by Vaithilingam et al. [41]. They found that Copilot did not necessarily reduce task duration, especially if the generated code was non-optimal, or increase task completion. Nevertheless, participants preferred using Copilot since it offered a good starting point for solving programming-related problems.

Sarkar et al. [34] gathered insights on user experiences with coding assistants by qualitative data analysis of blog posts. Whilst programmers found tools useful for boilerplate coding, challenges regarding effective prompting and checking and debugging generated, unfamiliar code remained. The authors took a programmer-centric approach to analyze the interaction with GenAI tools, yet the developer experience has not been validated with objective quantitative data.

Research that explicitly focuses on developers' well-being at work was conducted by Ngwenyama et al. [29]. They inspected the contribution of GenAI to work satisfaction in a questionnaire-based study. Their findings indicate that Copilot contributes to increased satisfaction, promotes a sense of flourishing at work, reduces frustration and increases perceived productivity.

Our research adds to the above-mentioned literature on developers' interaction behavior, experience, e.g. in terms of satisfaction, and productivity gains when interacting with GenAI. We seek to take a developer-centered mixed-methods analysis approach by combining quantitative and qualitative subjective with behavioral data gathered from professional developers within their natural work environment. As Bird et al. [4] state: "The challenge [is to create] the right user experience such that the developer is helped more than hindered".

## 4 STUDY

### 4.1 Procedure

Professional software developers from SAP participated in the study, which spanned four days (Figure 1). On the first and last day, they partook in controlled sessions. Between the controlled sessions, they worked for three days in an uncontrolled setting. During the uncontrolled phase, they continued their normal everyday work and documented the tasks they were working on. For each task, the start and end time were noted, perceived cognitive load and productivity were rated on Likert scales, and the use (yes/no) and perceived helpfulness of GenAI were indicated.

Multiple data sources were considered in this study: Subjective participant data was gathered via questionnaires, behavioral data was collected via screen recordings, keyboard use and mouse movement, and physiological activity was measured via a wristband. The evaluation focus of this paper is on the subjective questionnaire data and the screen recordings.

**First controlled session.** At the beginning of the first controlled session, each participant was informed about the study procedure, data collection, and data privacy, and then they provided their informed consent to participate in the study under these conditions. The participant then worked on a provided laptop whilst behavioral data via screen recording [31], keyboard use and mouse movement [3] and physiological data from a wristband were gathered.

1. First, a pre-questionnaire with questions on the work environment and experience [37], GenAI usage and satisfaction [20, 24], work satisfaction [33, 37], developer experience and flow state [11, 25, 39] and a 50-item International Personality Item Pool (IPIP) [14, 15] personality test was filled in (Figure 1, Pre-Questionnaire).
2. Then followed a cognitive task, during which the participant needed to remember alphabetical letters from n positions back (difficulty levels: n=1,2,3) (Figure 1, Cognitive n-back tasks). After the 3-back task, a startle event in the form of a honking sound occurred (Figure 1, Cognitive n-back tasks with a startle event), followed by a five-minute 4-7-8 breathing meditation session [18] (Figure 1, 4-7-8 breathing meditation). The cognitive task, together with the startle event, was conducted to analyze the physiological data, which is not the focus of this paper.

3. Then, the participant completed three baseline coding tasks in Java (one without and two with GitHub Copilot) to become familiar with the Visual Studio Code Integrated Development Environment (VS Code IDE), the GitHub Copilot integration (from here on only referred to as Copilot), and the task structure based on the HumanEval-X dataset [44] (Figure 1, 3 baseline coding tasks). For all coding-related tasks in this study, code metrics were calculated prior to the study to select tasks with comparable cognitive load [7].
4. After the baseline tasks, followed the main phase of the session during which the participant completed six randomized *tasks: coding, debugging, code documentation, writing unit tests, summary and brainstorming tasks* (Figure 1, 6 main tasks). Half of the participants did not use Copilot during the tasks of the first controlled session, whilst the other half did (Figure 1, A/B study design). Between tasks, the participant evaluated the perceived workload of the worked-on task by rating six 21-point Likert-scaled questions from the NASA-Task Load Index (NASA-TLX) [17] (Figure 1, NASA-TLX questionnaire). A one-minute relaxation video showing a nature scene followed the NASA-TLX questionnaire to provide the participant with a rest before the next task (Figure 1, Relaxation video). If the participant belonged to group A that did not use Copilot to work on the main tasks, the Copilot evaluation questionnaire (based on the SPACE framework [12] and adapted to evaluate specifically Copilot) was filled in after the baseline tasks. If the participant was part of group B, the Copilot evaluation questionnaire was filled in after the main tasks at the end of the first controlled session (Figure 1, Copilot evaluation questionnaire).

**Second controlled session.** The second controlled session on the last day of the study was structured similarly to the first session, with first a five-minute 4-7-8 breathing meditation exercise and then six tasks (same task categories as during the first controlled session) – this time performed by all participants with the help of Copilot (Figure 1, A/B study design) – and a NASA-TLX questionnaire with a relaxation video between each task. At the end of the second session, a post-questionnaire [24, 37] with questions regarding the participant's interaction with GenAI and an evaluation of the study procedure and setup was filled in (Figure 1, Post-Questionnaire). After the study, each participant received a 50$ Amazon voucher as compensation for their time and effort for participating via email.

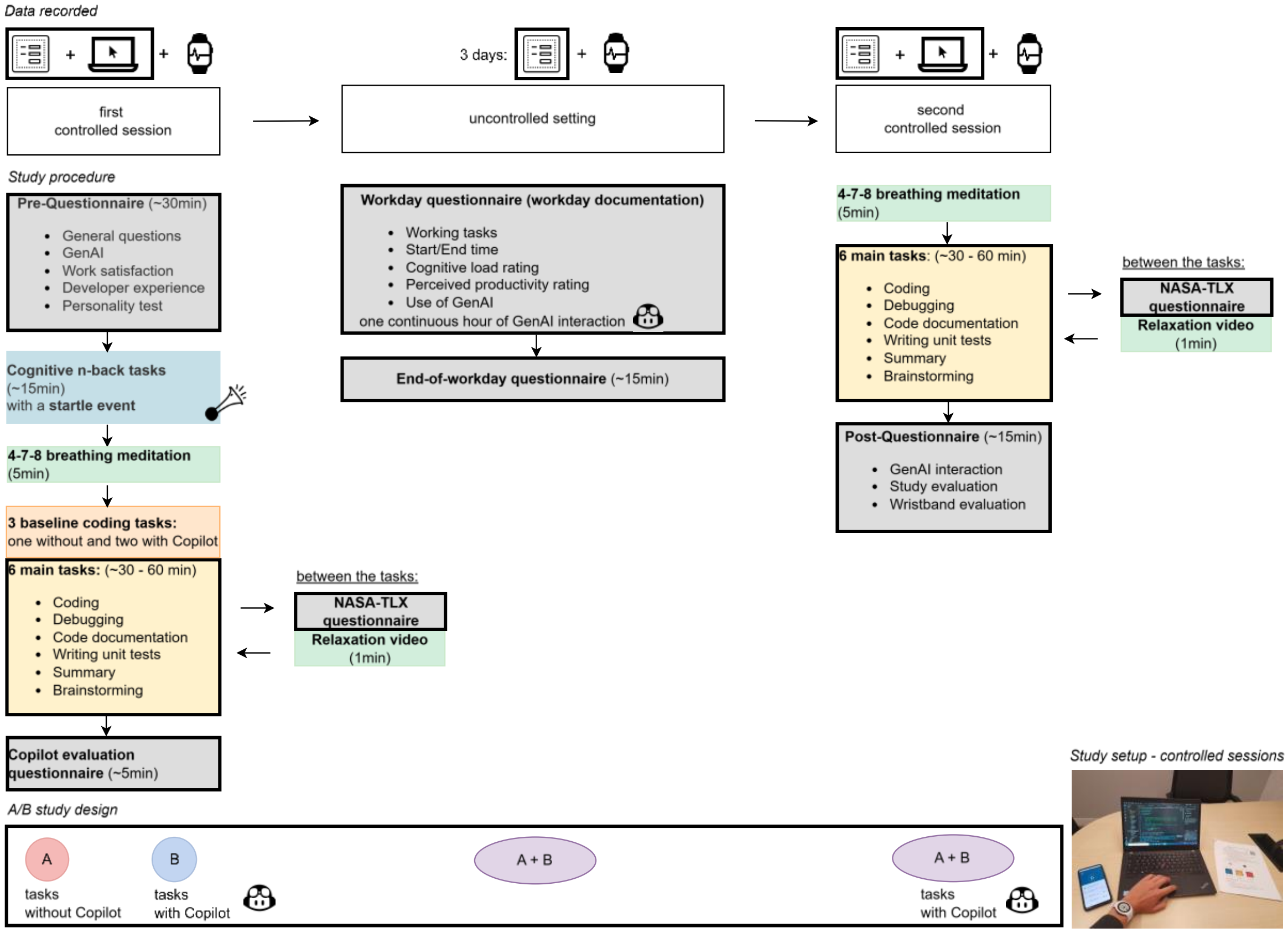

Figure 1: The study procedure, study setup and A/B study design are depicted. The recorded data during the study were subjective data from questionnaires, behavioral data from screen recordings and physiological data from a wristband. The sections with thick black surroundings indicate the evaluation focus of this paper. This figure was taken and adapted from [7].

### 4.2 Participants

The participant cohort consisted of twenty-two SAP employees from two firm sites in California, USA. The participant group included 12 software developers/engineers, 8 senior software developers/engineers, one senior quality specialist, and one principal software architect, all with an average of 5 to 10 years of experience in IT/programming. Java experience was a requirement for participation, and all participants reported using Java, with Python and JavaScript also commonly used. According to self-ratings, the participants were highly proficient in Java and also proficient in using GenAI. GitHub Copilot and ChatGPT were the most commonly used AI tools at work, whilst ChatGPT and Gemini were preferred outside of the work context. On average, participants had been using GenAI for 6-12 months and had experience for at least 1-6 months. For one participant, only questionnaire data from the controlled sessions could be analyzed due to non-adherence to the remaining study procedure. No age or gender information was collected for data privacy reasons. Participation in this study was voluntary, and the design and procedure of the study were reviewed and approved by the ethics committee of the University of Potsdam.

### 4.3 GitHub Copilot

GitHub Copilot is one of several GenAI tools that can assist developers by generating, completing and modifying programming code based on the context of the codebase and natural language prompts. It is integrated into IDEs like VS Code and can function as an "AI pair programmer" for developers. Participants could flexibly switch between different interaction types (in-code suggestions, chat, in-line chat), and modes (ask, edit, agent) and models.

The three primary interaction types differ in where code suggestions appear, the scope of the codebase considered for the output, their response latency, and the expected use:

- **In-code suggestions**: Suggestions are generated at the cursor position with low latency based on local code context. This supports fluid, autocompletion-like workflows.
- **In-line chat**: Localized questions or code modifications are enabled directly within the source file. The considered context is limited to the surrounding lines of code.
- **Chat**: A continuous, dialogue-style interaction takes place in a separate window. A broader context such as an entire file or multiple files is considered. Output in this mode is optimized for creativity and explanation.

GPT-4o was set as the default model and "ask" as the default chat mode for the controlled sessions. During the uncontrolled period, participants were free to choose a GenAI tool they wanted to interact with. Consistent with our research goal, the study focuses on analyzing the developers' interaction with GenAI tools, rather than evaluating specific tools or model performances.

### 4.4 Analysis methodology

#### *4.4.1 Data labeling*

Approximately 66 hours of participants' screen recordings from the two controlled sessions were labeled according to session phase (e.g. start of recording, pre-questionnaire start, pre-questionnaire end etc.) and main tasks (e.g. debugging start, debussing end, NASA-TLX, coding start etc.). To capture developers' interactions with GenAI during the controlled sessions, we furthermore manually annotated the main tasks of the controlled sessions. Annotations were applied per participant and task, capturing temporal information (session day and task order), task outcomes (duration and completion), and interaction characteristics with Copilot (with or without Copilot use, number of in-code suggestions and number of chat prompts). Interactions with Copilot's in-line chat were treated as chat interactions. To ensure consistency, ambiguous cases during annotation were reviewed by two further authors of this paper.

#### *4.4.2 Data categorization*

The 445 working tasks documented by the participants in the workday questionnaire (Figure 1) were categorized into three general categories: development-heavy, collaboration-heavy and other activities. The categorization and groupings of the tasks (Table 1) is based on Meyer et al. [23] and adaptations from Brandebusemeyer et al. [6]. Minor adjustments to the categorization schema were made in this paper, e.g. by assigning pair-programming as a development-heavy activity instead of as a collaboration-heavy activity [6]. Three authors were involved in the manual categorization of the data to reduce individual bias in this partly interpretive process.

Table 1 Working task categories based on Meyer et al. [23] and Brandebusemeyer et al. [6].

| General task categories | Groupings of participants' documented tasks |
|---|---|
| Development-heavy activities | Coding<br>Debugging<br>Testing<br>Specs/requirements<br>Reviewing<br>Documentation<br>Pair-programming |
| Collaboration-heavy activities | Meetings (also discussions)<br>Messaging (emails, slack etc.)<br>Helping/mentoring<br>Networking |
| Other activities | Honing skills/continuous learning<br>Admin tasks<br>Multitasking<br>Breaks (lunch, socializing, walking etc.)<br>Various (planning etc.) |

#### *4.4.3 Statistics*

We focus on analyzing the questionnaires from the controlled sessions and the workday questionnaire from the uncontrolled study period, and the analysis of the behavioral data gathered from the screen recordings of the controlled sessions. This way, subjective experiences and behavioral interaction with GenAI contribute to a holistic evaluation.

The Likert-scaled questions were evaluated via descriptive statistics, Kendall's tau correlation analysis or linear mixed-effects models. Models were constructed when cognitive load and productivity ratings during the three different task categories were compared. Open-ended responses were analyzed via qualitative content analysis according to the guidelines by Mayring & Fenzl [22].

To predict task completion, duration, and perceived workload during the controlled sessions, (generalized) linear mixed-effects models were used. In a previous analysis [5] we found that neither the two controlled session days nor the two groups from our A/B study design (Figure 1, A/B study design) differed significantly, so controlled sessions and groups are analyzed together.

## 5 RESULTS

### 5.1 Developers' interaction experience with AI before the study

We first want to get an impression of the participants' sentiment toward AI/GenAI before interacting with it in the study. The answers to the pre-questionnaire are evaluated in the following.

#### *5.1.1 Sentiment on AI use during work*

We first examined the participants' various attitudes towards using AI at work with a 5-point Likert scale (1=strongly disagree, 5=strongly agree) (Figure 2). AI was rated most helpful for monotonous tasks (M=4.50), with all participants agreeing or strongly agreeing. Participants also generally agreed that AI increases their productivity (M=4.32), makes their work less strenuous (M=4.05), and that GenAI tools are helpful pair programmers (M=4.0). AI was perceived as supportive even during enjoyable tasks (M = 3.64), though participants were more neutral regarding its usefulness for creative tasks (M = 3.29). Participants tended to agree that AI helps make their overwhelming workloads more manageable (M = 3.57), reduces mental effort compared to working without AI (M = 3.59), and improves their work quality (M = 3.59), though five participants disagreed on the latter. The greatest disagreement (10 participants) arose with the statement "AI makes me feel more fulfilled in my work" (M=2.91). Overall, attitudes toward AI at work were positive to neutral. During the analysis of the controlled sessions, we are going to examine tasks that can be considered monotonous, enjoyable or creative, and therefore these categories are already marked in bold in Figure 2.

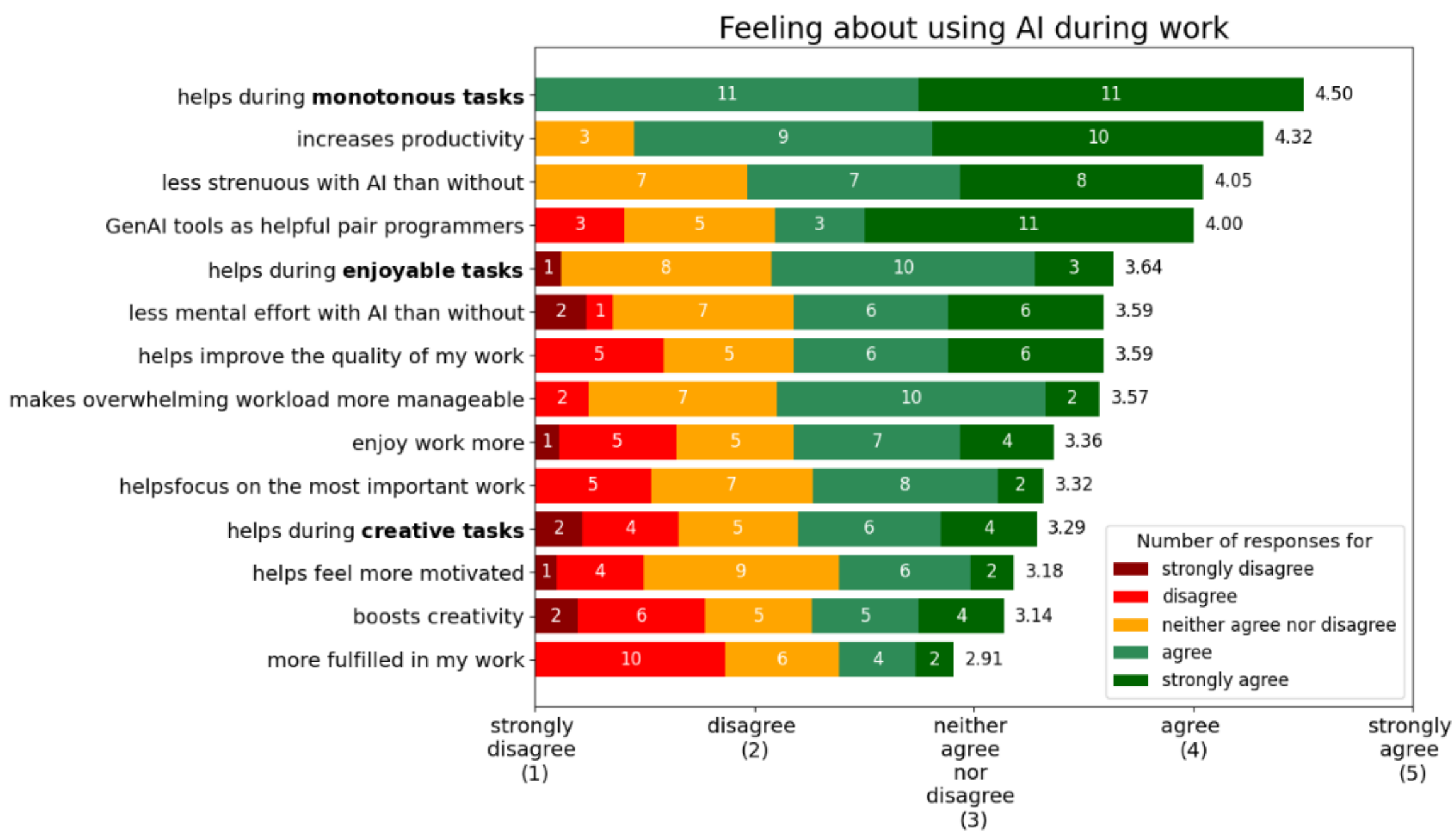

Figure 2: The figure illustrates participants' feelings about using AI during work and their level of agreement with the statements on the y-axis. The numbers within the bars indicate the number of participants with a specific response. The average response across participants is shown at the end of each bar. Two participants selected "not applicable" and are therefore not contained in this figure.

### 5.1.2 *GenAI use across software engineering tasks*

We now focus on understanding the impact of GenAI on developers' working tasks. First, we analyzed what a regular work week looks like for the developers (Figure 3, A), to then examine how frequently GenAI was used for software engineering tasks (Figure 3, B).

To gain insight into the participants' typical work week, they were asked to estimate the number of hours spent on a given list of working tasks regardless of the AI usage. The tasks and the average percentage of time spent on them per week across 21 participants are depicted in Figure 3, A. One participant needed to be excluded from the analysis, since working tasks were not documented. Most of the time during a regular week, developers spend on writing code (29.04%), meetings (17.49%), debugging (13.91%), and running tests (10.31%).

Then, GenAI use across different task categories was analyzed (Figure 3, B). The participants rated how frequently they use GenAI for common software engineering tasks on a 4-point Likert scale (1=not at all, 4=to a great extent). GenAI was most frequently used for writing code (M=3.18), writing and running tests (M=3.14) and debugging (M=3.0). In contrast, it was used less for creative tasks such as writing text (M=2.55), brainstorming (M=2.45) or summarizing (M=1.73). These usage patterns align with the findings on participants' perceived helpfulness of AI at work (Figure 2). Additionally, the three tasks for which GenAI is most frequently used are also the ones that consume the most time during a regular work week for developers. GenAI is therefore used for the most time-consuming coding-related working tasks. The tasks of coding, debugging, documentation, testing, summarizing, and brainstorming (printed bold in Figure 3, B) are further examined when analyzing the controlled sessions of this study.

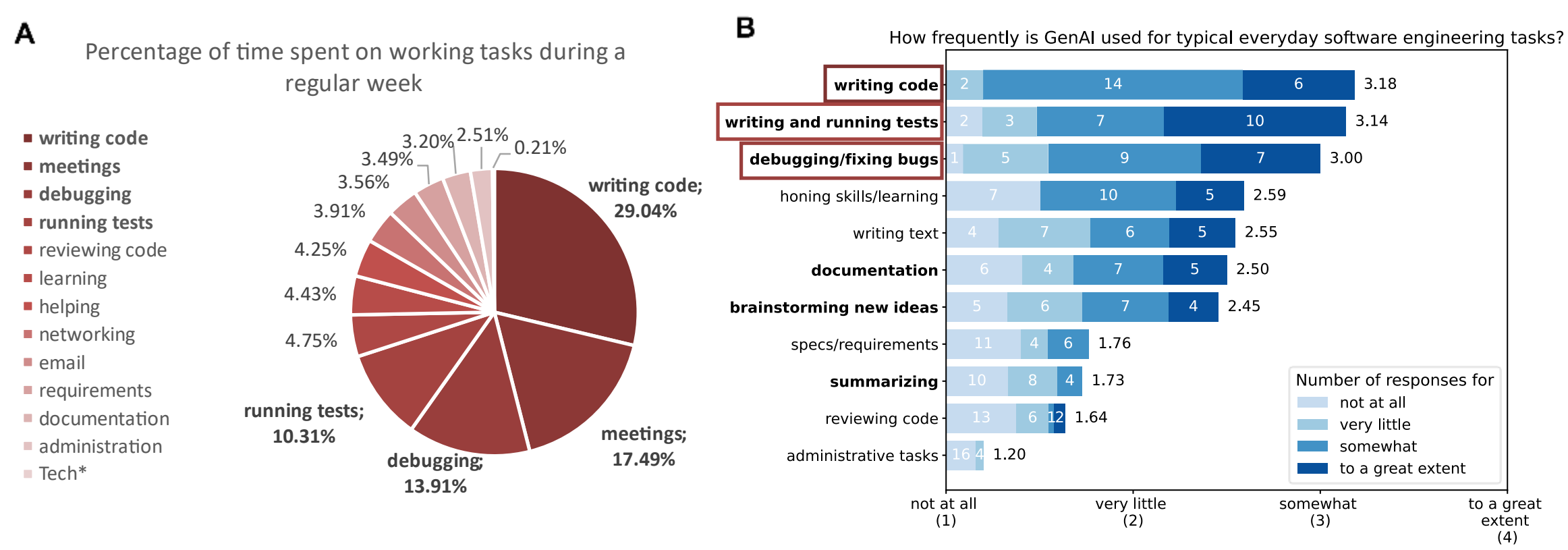

Figure 3: **A:** The average percentage of time that the participants spend on working tasks during a regular week is depicted. The task marked with * is an additional task category added by a participant. **B:** The participants rated how frequently they used GenAI tools (4-point Likert scale on the x-axis) for the tasks displayed on the y-axis. The lengths of the bars and the numbers at the ends indicate the average frequency of GenAI use according to the x-axis. In three cases, participants selected "not applicable", so the answers did not contribute to this figure.

#### *5.1.3 Satisfaction with GenAI*

On being explicitly asked how satisfied they are overall with the interaction with LLMs / GenAI tools, most participants reported being either satisfied (n=11) or very satisfied (n=5) (72.73%) with none expressing dissatisfaction. Six participants were neither satisfied nor dissatisfied with the interaction so far (Figure 4).

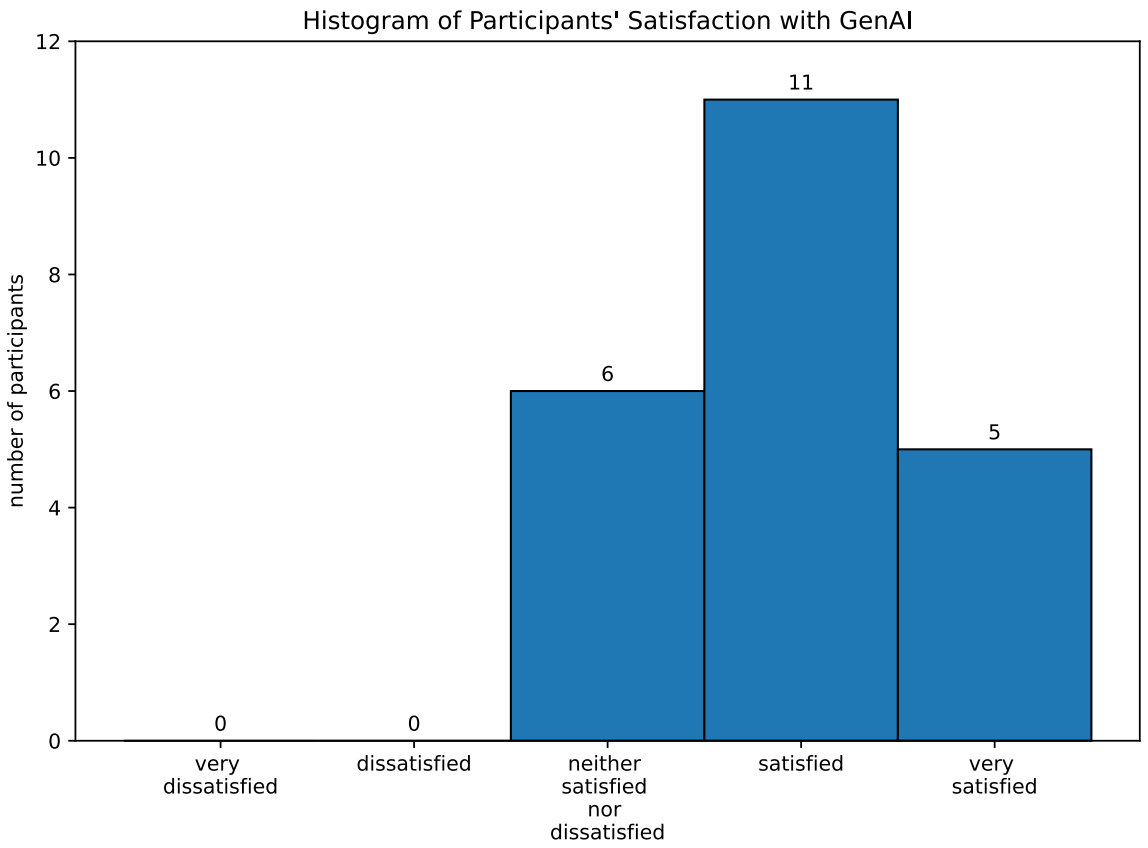

Figure 4: Histogram of participants' satisfaction with GenAI.

To understand what drives satisfaction and what the sources of dissatisfaction are, open-ended responses were categorized into three areas: tool/model output quality, tool/model interaction quality, and work efficiency (Table 2).

Participants expressed satisfaction with GenAI output quality, particularly for unit testing, debugging, and completing time-consuming or tedious tasks that do not require thought or innovation. These results align with their perception of AI's usefulness for monotonous tasks (Figure 2 and Figure 3). However, dissatisfaction arose from inaccuracies when dealing with more complex topics and hallucinations.

Regarding interaction quality, six participants valued GenAI for providing them with new ideas and fresh perspectives on a problem, especially in the absence of prior knowledge. Although valuing GenAI's creativity, explicit use of it, for e.g. brainstorming, is still underutilized (Figure 2 and Figure 3). Challenges with the interaction include prompt sensitivity – requiring repeated prompt adjustments for the desired output – and the need for constant human verification.

In terms of work efficiency, all but one of the mentioned aspects by the participants are positive. Participants are satisfied with the efficiency of writing (boilerplate) code, unit tests and code documentation. Again, mostly monotonous tasks are mentioned in the context of efficiency, indicating that participants value support from GenAI regarding monotonous tasks. One participant indicated that slow model response times negatively affected efficiency for simple tasks.

Table 2: Open-ended questions regarding satisfying and dissatisfying aspects of GenAI were categorized. The numbers n represent the number of participants mentioning an aspect.

| Categories | Satisfied | n | Dissatisfied | n |
|---|---|---|---|---|
| *Level 1* | *Level 2* | | *Level 2* | |
| **Tool /model output quality** | • writing unit tests<br>• accurate task completion and information for simple or time-consuming/tedious tasks without thought or innovation needed<br>• debugging | **3**<br>**3**<br>2<br>1 | • inaccuracy in results (when dealing with complex logic/topics)<br>• hallucinations<br>• (outdated) internal models<br>• business domain knowledge/ context missing | **7**<br>3<br>2<br>2 |

| Categories | Satisfied | n | Dissatisfied | n |
|---|---|---|---|---|
| *Level 1* | *Level 2* | | *Level 2* | |
| | • helps follow best coding practices<br>• suggests prompts<br>• standardized solution | 1<br>1 | • verbose code that is hard to review | 1 |
| **Tool/model interaction quality** | • new ideas/suggestions/ hints/input/options for different solutions and new ways to think/go about a problem - starting point if no prior knowledge is present<br>• chat for threaded discussion/feedback option<br>• tool/model integration/interface, e.g. context from other coding tabs, IntelliJ integration<br>• better interpretation of what one is trying to communicate<br>• information retrieval from large data sources<br>• tailored solutions via interface instead of web search<br>• learning a new programming language | **6**<br>2<br>2<br>1<br>1<br>1<br>1 | • prompt sensitivity - desired results sometimes require repeated prompts<br>• human verification needed<br>• fails to acknowledge when it does not have to answer<br>• over-confidence in output correctness<br>• response quality not always evident<br>• amount of context needed to be provided beforehand<br>• internal UI subpar the native UIs of the model providers – makes many models unusable | **2**<br>**2**<br>1<br>1<br>1<br>1<br>1 |
| **Work efficiency** | • efficient (boilerplate) coding<br>• efficient writing of unit tests<br>• time reduction for code documentation<br>• quick solutions (for issues)<br>• more code produced (save time on continuous brainstorming)<br>• reduces research time | **4**<br>3<br>2<br>2<br>1<br>1 | • speed as the determining factor for liking the interaction with the model – often for simple tasks, waiting for the model takes too long | 1 |

Table 3 shows the number of participants who mentioned at least one satisfying or dissatisfying aspect regarding a category. Satisfaction and dissatisfaction were balanced for output and interaction quality, while all but one valued the work efficiency with GenAI.

Table 3: The number of participants who mentioned aspects in each of the three categories is summarized in the table.

| Categories | Satisfied | Dissatisfied |
|---|---|---|
| **Tool /model output quality** | 10 | 10 |
| **Tool/model interaction quality** | 10 | 9 |
| **Work efficiency** | 9 | 1 |

In sum, in the pre-questionnaire of this study, participants reported satisfaction with GenAI for repetitive and structured tasks such as testing, debugging and boilerplate coding, aligning with their frequent use and perceived usefulness during monotonous tasks. While GenAI was appreciated for generating new ideas, creative use cases like brainstorming remained less common. Dissatisfaction centered on output inaccuracies and prompt sensitivity whilst benefits in work efficiency and productivity were seen.

### 5.2 Developers' interaction behavior with GenAI during the controlled sessions of the study

To identify developers' preferred way of interacting with GitHub Copilot, i.e., preferred interaction type, the following analyses focus on screen recording data from the two controlled sessions and the Copilot evaluation questionnaire.

#### *5.2.1 Dominant interaction type during controlled sessions*

Figure 5 shows that in-code suggestions were the dominant interaction type whilst working on tasks during the controlled sessions. 14 out of 20 participants used in-code suggestions more than 50% of the time compared to chat prompts.

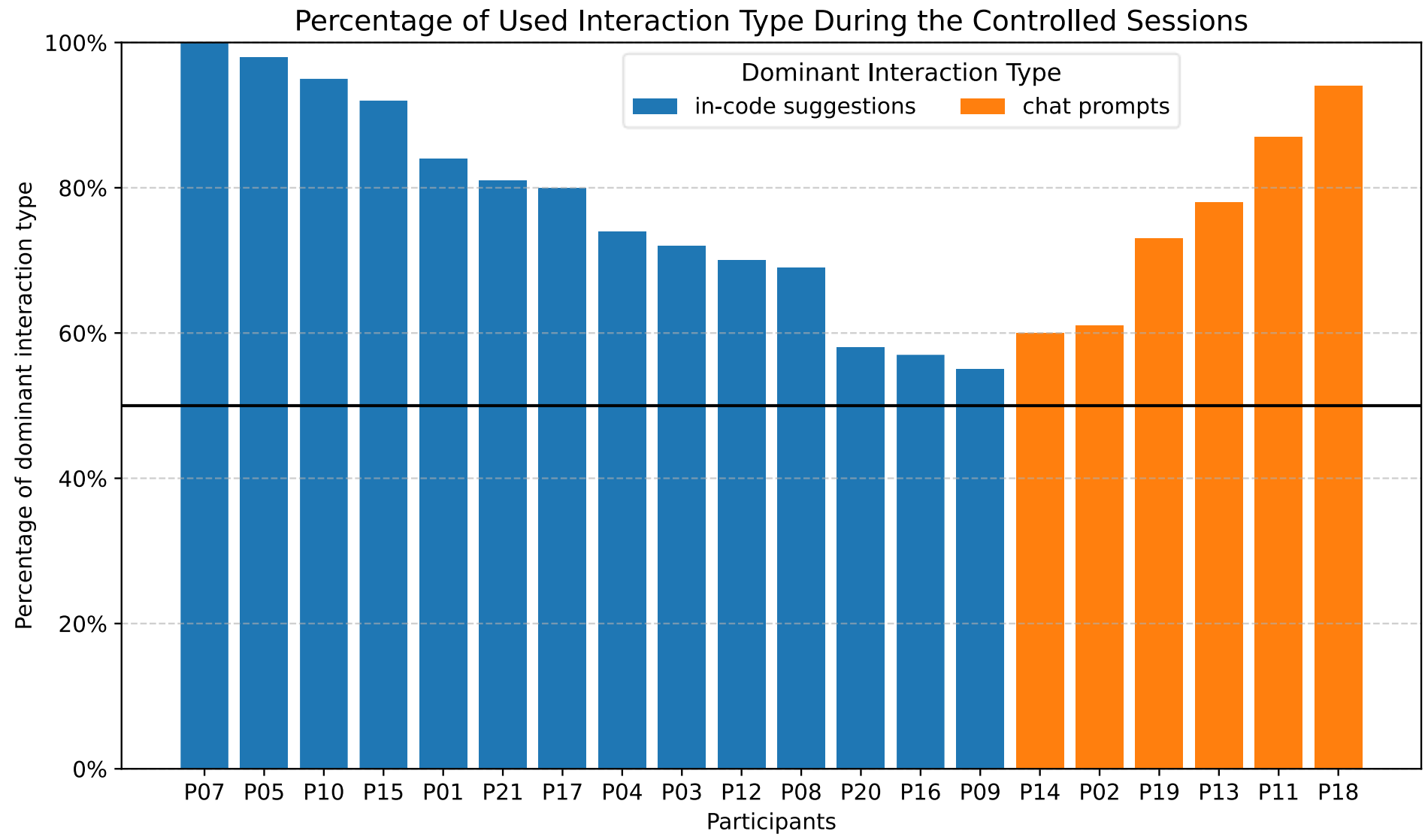

Figure 5: The percentage of the participants' most used interaction type with GenAI during the tasks of the controlled sessions is depicted. One participant was excluded, since the data from the first controlled session is missing.

#### *5.2.2 Copilot evaluation after first use in this study*

After their initial interaction with Copilot during the first controlled session, participants completed the Copilot evaluation questionnaire. Group A completed it after the three baseline tasks (two of which involved using Copilot), while Group B completed it after working on the main tasks with Copilot as well.

Participants first indicated which interaction type they had used (in-code suggestions, chat, or both). If both was selected, they answered the same questions once for in-code suggestions and once for chat prompts. The questions can be assigned to one of the three categories: output quality, interaction quality or work efficiency (Figure 6).

Participants rated the output quality of both interaction types as providing relevant suggestions and being easy to understand. Output quality is therefore evaluated as comparable between the interaction types.

In terms of interaction quality, both interaction types were seen as easy to use and recommendable as pair programmers. Participants also felt less frustrated and less stressed when programming with Copilot. Interaction types diverged regarding satisfaction: participants using in-code suggestions on average agreed that they feel more satisfied when interacting with Copilot, whilst participants on average neither agreed nor disagreed for chat use ($M_{inCode}$=3.72, $M_{chat}$=3.4).

Participants' agreement to the questions on work efficiency, whether they feel more productive when using Copilot, was higher for in-code suggestions (M=4.33) than chat interaction (M=3.8). Both interaction types were seen as saving time ($M_{inCode}$=4.5, $M_{chat}$=4.3).

Overall, while both interaction types with Copilot were positively evaluated, in-code suggestions were rated more favorably for satisfaction and productivity. However, the preference for in-code suggestions seems to be task-dependent. Participants who used both interaction types during the tasks were asked in an open-ended question whether they preferred in-code suggestions or chat. The consensus was that in-code suggestions were preferred for coding tasks that are less complex and require small changes, whilst chat is preferred for debugging and non-coding situations where a broader context needs to be considered and explanations for the output are favored.

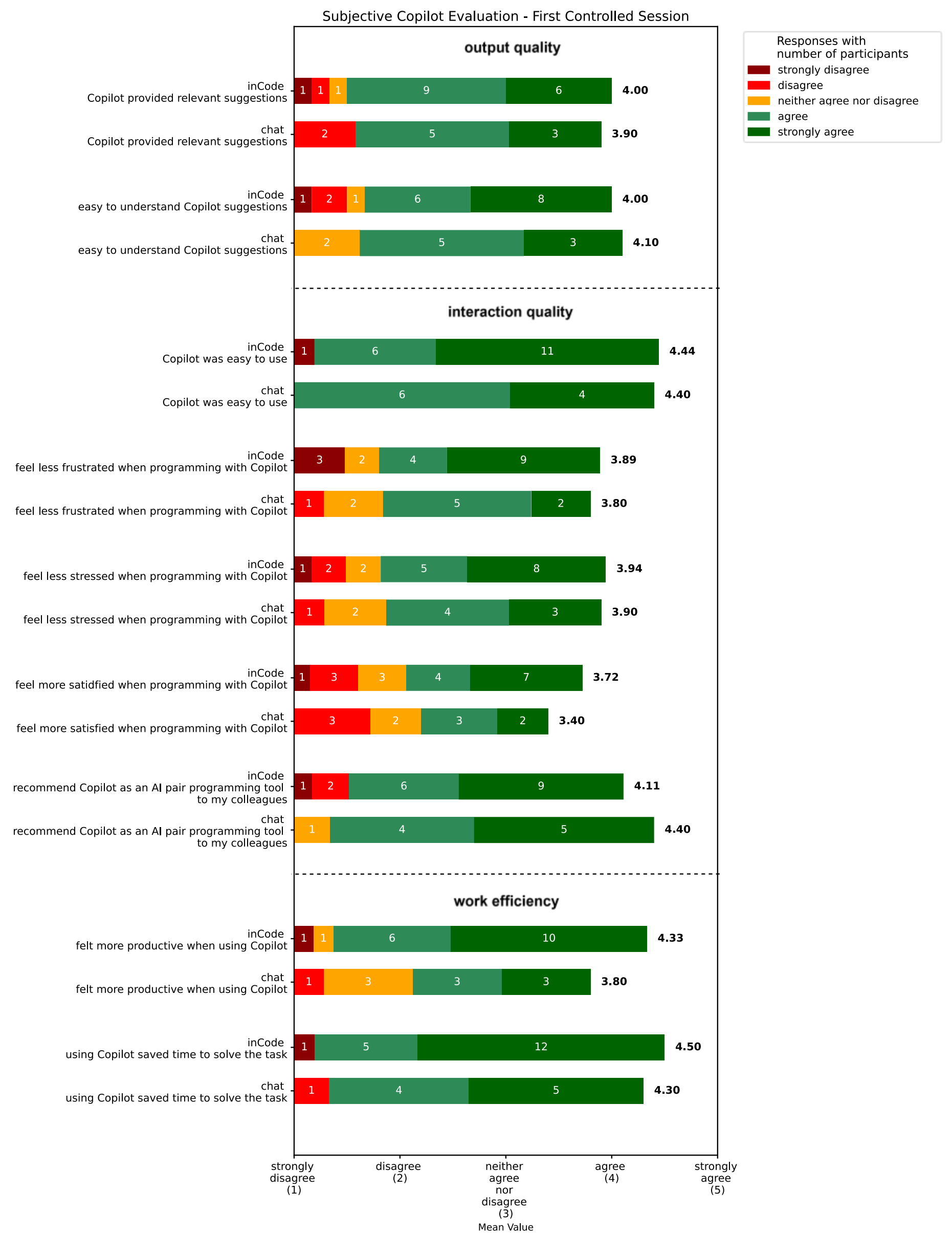

Figure 6: Subjective Copilot evaluation of the participants during the first controlled session.

Developers preferred in-code suggestions over chat during their first interaction with Copilot in this study. In-code suggestions were used more often and rated higher in satisfaction and productivity. However, preferences vary by task type and complexity: in-code suggestions were favored for simple code changes, while chat was preferred for debugging and tasks requiring broader context or explanations.

### 5.3 Impact of Copilot on efficiency, accuracy, and perceived workload

In our previous work [7], the impact of Copilot on task efficiency (task duration), accuracy (task completion) and perceived workload during the controlled sessions of the study was analyzed in detail based on the screen recordings and the NASA-TLX scores. Since we would like to interpret these previous results with the subjective quantitative and qualitative evaluations from the questionnaires during the controlled and uncontrolled study phases, we summarize the findings detailed in [7] in the following.

#### *5.3.1 Efficiency (task duration)*

Copilot usage significantly affected task duration. Using either in-code suggestions or chat prompts individually led to significantly shorter task durations compared to not using Copilot (Figure 7). However, combining both interaction types did not yield additional efficiency benefits and resulted in task durations comparable to the no-Copilot condition (Figure 7). Interaction intensity further moderated these effects: moderate use of either in-code suggestions or chat prompts improved efficiency, while excessive interactions, particularly frequent chat prompts or combined interaction modes, introduced overhead that diminished or reversed time savings. Task categories also influenced efficiency, with debugging and summary tasks taking significantly longer than coding tasks. Higher perceived workload was associated with longer task durations.

#### *5.3.2 Accuracy (task completion)*

Task completion rates were influenced by both Copilot interaction type and task category. Using only chat-based prompts significantly increased the likelihood of task completion compared to no Copilot usage, while in-code suggestions alone or combined usage showed no significant effect. The task category served as a predictor of task completion: coding and brainstorming tasks exhibited the highest success rates, whereas debugging, testing, documentation, and summary tasks were significantly less likely to be completed successfully. These results suggest that chat-based GenAI support is particularly beneficial for tasks that require higher-level reasoning or explanation.

#### *5.3.3 Perceived workload*

The Copilot interaction type had an impact on perceived workload. Using either in-code suggestions or chat prompts alone significantly reduced perceived workload compared to not using Copilot. In contrast, combining both interaction types resulted in workload levels comparable to the no-Copilot condition and significantly higher workload than using in-code suggestions alone (Figure 7), suggesting that switching between interaction modes increases cognitive overhead. While task categories influenced overall workload ratings, no individual task type differed significantly from coding, supporting the study’s design goal of comparable cognitive load across tasks.

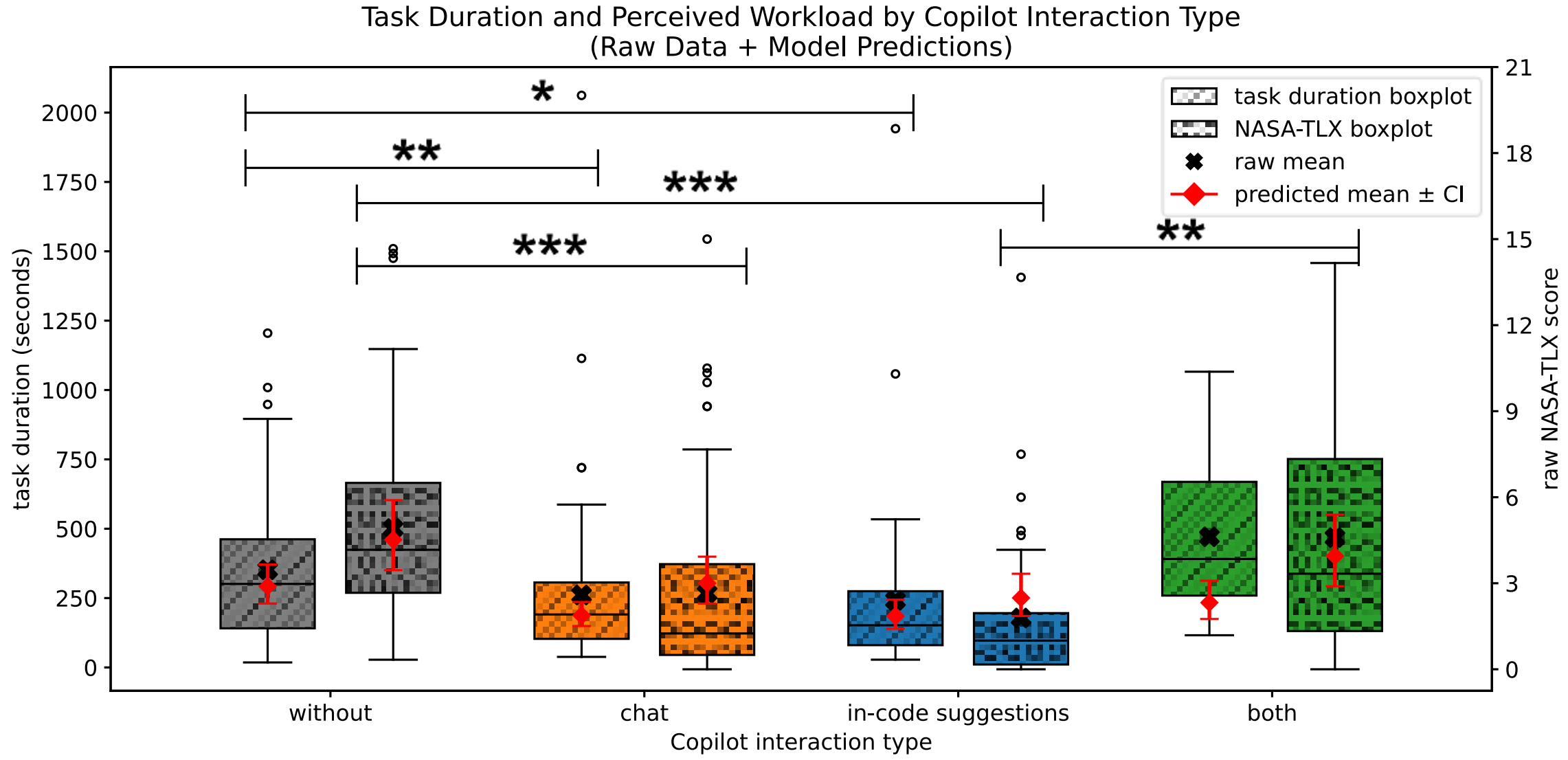

Figure 7: The raw data and model predictions of task durations and perceived workload (raw NASA-TLX scores) grouped by Copilot interaction type are depicted. Asterisks denote statistical significance (* $p \leq 0.05$, ** $p \leq 0.01$, *** $p \leq 0.001$). Adapted from Figure 3 and Figure 6 from the Paper [7].

Use of either in-code or chat suggestions improves efficiency and reduces perceived workload, while combined use lessens these benefits. Excessive use of interaction types reduces efficiency benefits. Only chat use boosts task completion in this study. Task categories significantly affect efficiency and accuracy in this study.

### 5.4 Perceived cognitive load and productivity during everyday working tasks

#### *5.4.1 Descriptive analysis of the everyday working tasks*

During the uncontrolled period of the study (Figure 1, uncontrolled setting), the developers continued their everyday work whilst documenting their working tasks. Each working task was later categorized into development-heavy, collaboration-heavy and other tasks, as described in Section 4.4.2. For each task, participants documented whether they used AI and whether they found the interaction with AI helpful.

In total, 445 work tasks were documented by the developers. 41.6% of all tasks were development-heavy tasks, 28.1% were collaboration tasks, and 30.3% were other tasks. 120 tasks (27.4% of all documented tasks) involved AI interaction. Most AI was used for development-heavy tasks. Out of all development-heavy tasks, 54.6% involved AI use (Figure 8). When AI was used for development-heavy tasks, it was also perceived as helpful in 88.1% of cases (Figure 8). AI was used less for collaboration-heavy tasks and for other activities. 8% of all collaboration-heavy tasks and 8.1% of all other activities involved AI tools (Figure 8). However, if AI was used, it was almost always perceived as helpful for collaboration-heavy and other tasks (collaboration tasks: 90.9%; other tasks: 100%) (Figure 8). Differences in perceived helpfulness could be observed between participants: 12 developers rated AI as helpful every time they used it, eight developers found it sometimes helpful, one developer never found it helpful, and one did not use AI during the uncontrolled setting of this study [7].

In our previous analyses [7], tasks involving AI interaction were compared with those without. Significantly higher cognitive load, but also productivity was experienced when using AI than without [7]. In the following sections, a more nuanced analysis of participants' cognitive load and productivity ratings during the three working task categories (development-heavy, collaboration-heavy, and other) is conducted.

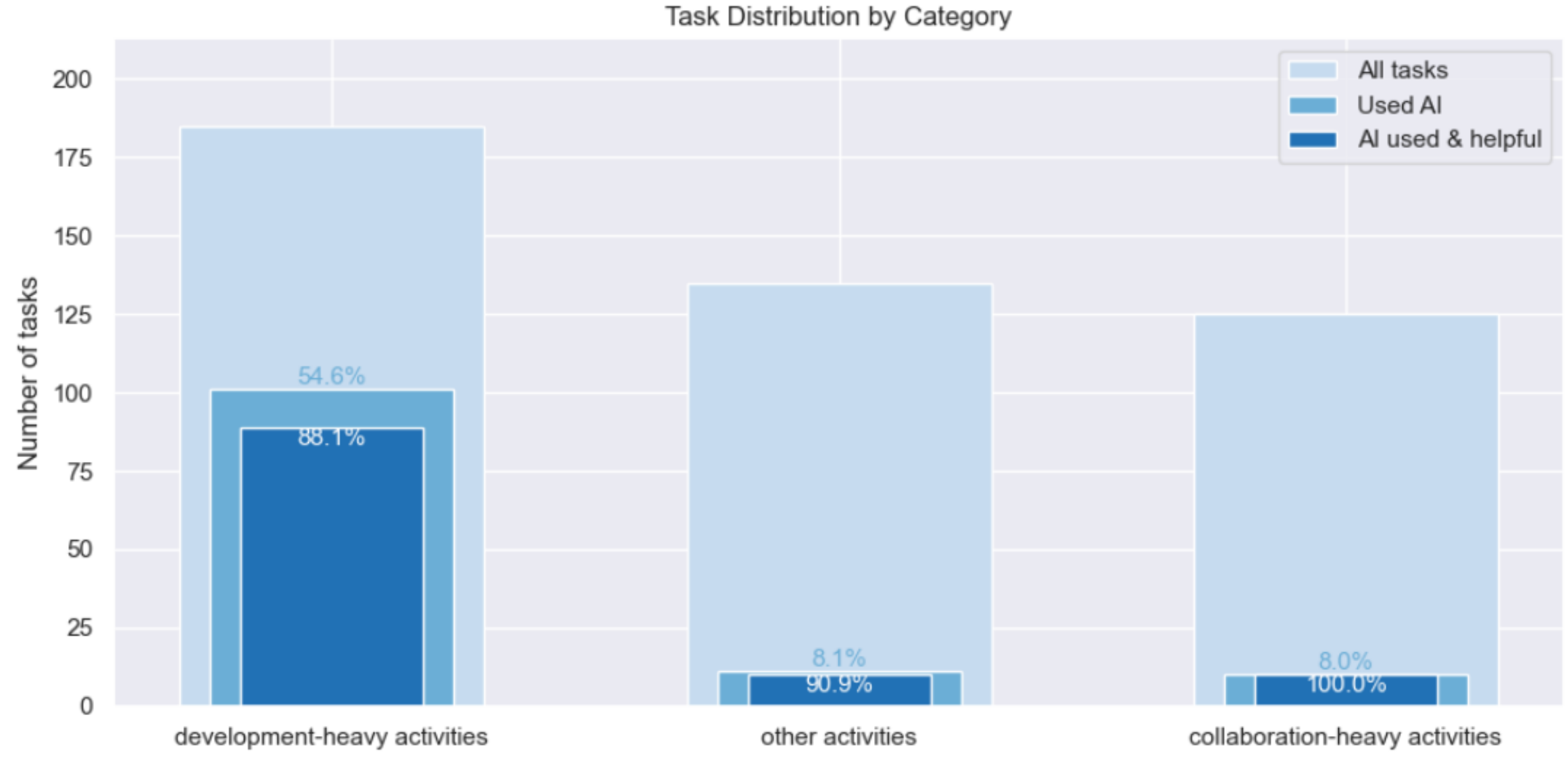

Figure 8: The distribution of working tasks per working task categories (development-heavy, collaboration-heavy and other) is depicted. The height of the light blue bars indicates the number of tasks per task category. The height of the middle blue bar shows the absolute number of tasks during which AI was used. The percentage in the middle blue denotes the amount of tasks with AI use out of all tasks of a category. The height of the dark blue bars denotes the absolute number of tasks in which AI was used and was perceived as helpful. The percentage in the dark blue bar indicates the amount of tasks in which AI use was perceived as helpful out of the tasks in which AI was used.

### *5.4.2 Difference between task categories*

To examine whether cognitive load and productivity ratings differ significantly between the three working task categories, linear mixed-effect models were constructed:

(1) cognitive load ~ categories + (1|participants)

(2) productivity ~ categories + (1|participants)

with the task categories as fixed effect and the participants as random effect.

As can be observed in Figure 9, the task categories differed significantly in cognitive load ratings ($F(2,419.44)=93.27$, $p<0.0001$). Estimated marginal means (EMMs) showed that development-heavy tasks were rated as the most cognitively demanding (EMM=4.57), collaboration-heavy tasks were intermediate (EMM=3.30), and other tasks were rated lowest (EMM=2.23). A pairwise post hoc test with Holm-Bonferroni adjustment showed that every task category differs highly significantly ($p<0.0001$) from the others in terms of cognitive load. Development-heavy tasks were rated as substantially more cognitively demanding than collaboration-heavy activities (difference=1.27, SE=0.17, $t(421)=7.38$, $p<0.0001$, $d=0.89$) and other activities (difference=2.34, SE=0.17, $t(424)=13.45$, $p<0.0001$, $d=1.63$). Cognitive load was also rated significantly higher for collaboration-heavy activities than for other activities (difference=1.07, SE=0.19, $t(419)=5.73$, $p<0.0001$, $d=0.75$). The results indicate that task categories affect participants' perceived cognitive load, with development-heavy tasks rated highest, followed by collaboration-heavy tasks, and then other tasks.

The task categories also differed significantly in productivity ratings ($F(2,415.66)=89.92$, $p<0.0001$). EMMs showed that participants rated their perceived productivity highest for development-heavy tasks (EMM=4.57), followed by collaboration-heavy tasks (EMM=3.30), and other tasks (EMM=2.25). Pairwise comparisons with Holm-Bonferroni adjustment indicated that all differences were highly statistically significant ($p<0.001$). Specifically, perceived productivity for development-heavy tasks was rated higher than for collaboration-heavy tasks (difference=1.27, SE=0.17, $t(417)=7.36$, $p<0.0001$, $d=0.89$) and other tasks (difference=2.32, SE=0.18, $t(421)=13.18$, $p<0.0001$, $d=1.62$). Collaboration-heavy tasks were also rated higher than other tasks (difference=1.05, SE=0.19, $t(415)=5.57$, $p<0.0001$, $d=0.73$). These results indicate that task category significantly affects perceived productivity, with the highest perceived productivity during development-heavy tasks.

Examining the random effect of the models (1) and (2) more closely, between-participant differences account for a variance of 0.45 in cognitive load and productivity ratings, while the residual variance (within-participant variability across tasks) was 2.05. The adjusted intraclass correlation coefficient (ICC) was 0.18, indicating that about 18% of the total variance in ratings is attributable to differences between participants, with the remaining 82% due to task-level variability. This means that participants differ moderately in their overall task ratings, but most of the variation in cognitive load and productivity ratings comes from differences between task categories. Therefore, these results justify including participants as a random effect in the linear mixed-effects models, as they explain part of the variance, while the main differences in cognitive load and productivity ratings can be attributable to the task categories.

In sum, the task categories significantly influenced both perceived cognitive load and productivity, with development-heavy tasks rated highest, collaboration-heavy tasks intermediate, and other tasks lowest. All pairwise differences between categories were statistically significant with medium-to-large to very large effect sizes, indicating that cognitive load and productivity ratings differed systematically across the task categories. Overall, tasks rated as more cognitively demanding were also rated higher in productivity.

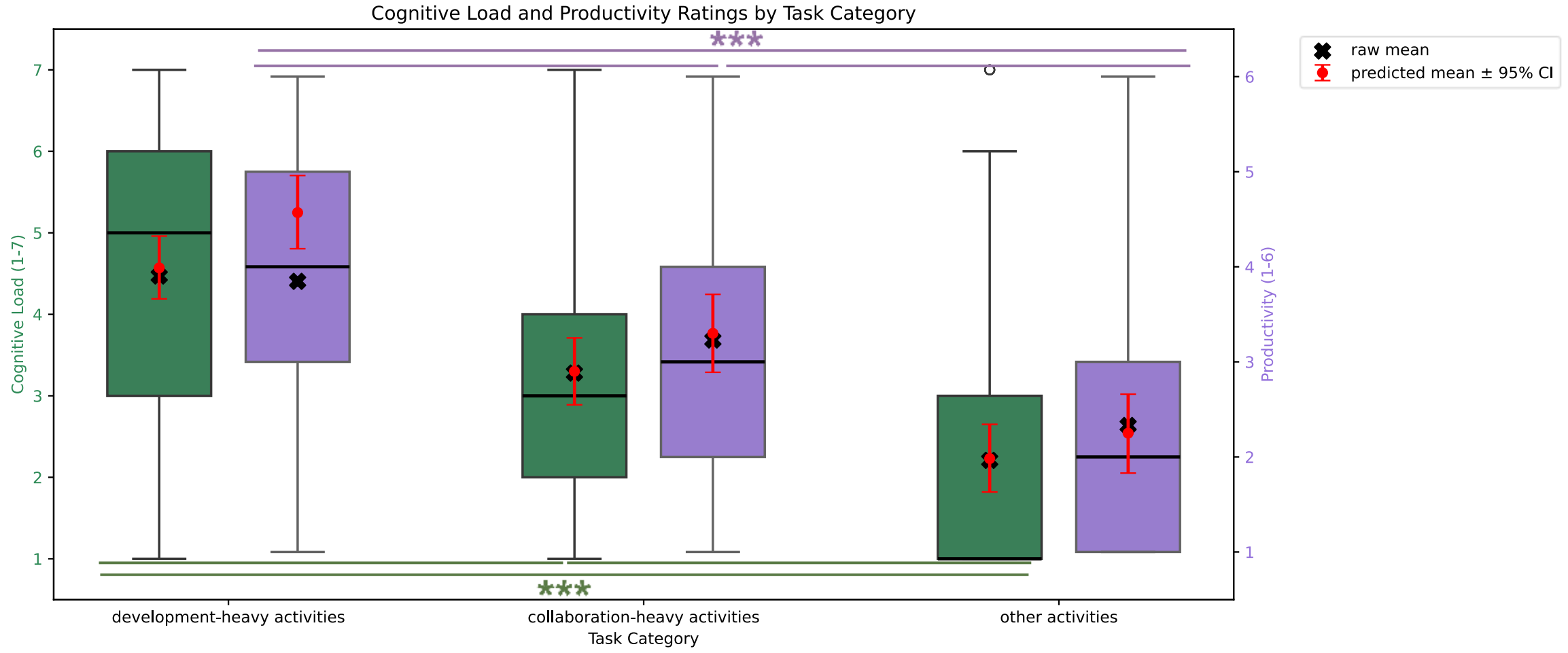

Figure 9: The cognitive load (green) and productivity ratings (purple) per task category are depicted. The lines below the boxplots indicate significant differences in terms of cognitive load between the task categories and the lines above the boxplots indicate significant differences in terms of productivity. The colored asterisks denote statistical significance for each line in the same color (*** p≤0.001), i.e. here all task categories differ significantly from each other in terms of cognitive load and productivity ratings.

#### *5.4.3 Effect of AI use on perceived cognitive load and productivity development-heavy tasks*

Figure 8 visualizes that AI tools were used more for development-heavy tasks than for collaboration-heavy and other tasks, which suggests examining AI use during development-heavy tasks more closely. Specifically, the question whether perceived cognitive load or productivity differ between development tasks with or without AI support is going to be considered in the following.

Linear mixed-effects models are constructed to answer this question:

(3) cognitive load ~ used_AI + (1|participants)

(4) productivity ~ used_AI + (1|participants)

with AI usage (yes vs no) as fixed effect and participants as random effect.

Cognitive load ratings differed significantly between development-heavy tasks with or without AI usage ($EMM_{yes}$=4.72, $EMM_{no}$=4.14, difference=0.58, SE=0.22, t(183)=2.67, p=0.008, d=0.34) (Figure 10). These results indicate that using AI in development-heavy tasks is associated with a small-to-moderate yet significant increase in perceived cognitive load.

Productivity ratings did not differ significantly between development-heavy tasks whether AI was used or not ($EMM_{yes}$=3.93, $EMM_{no}$=3.72, difference=0.22, SE=0.20, t(178)=1.10, p=0.275) (Figure 10). Unlike cognitive load, using AI in development-heavy tasks is not associated with a significant change in perceived productivity.

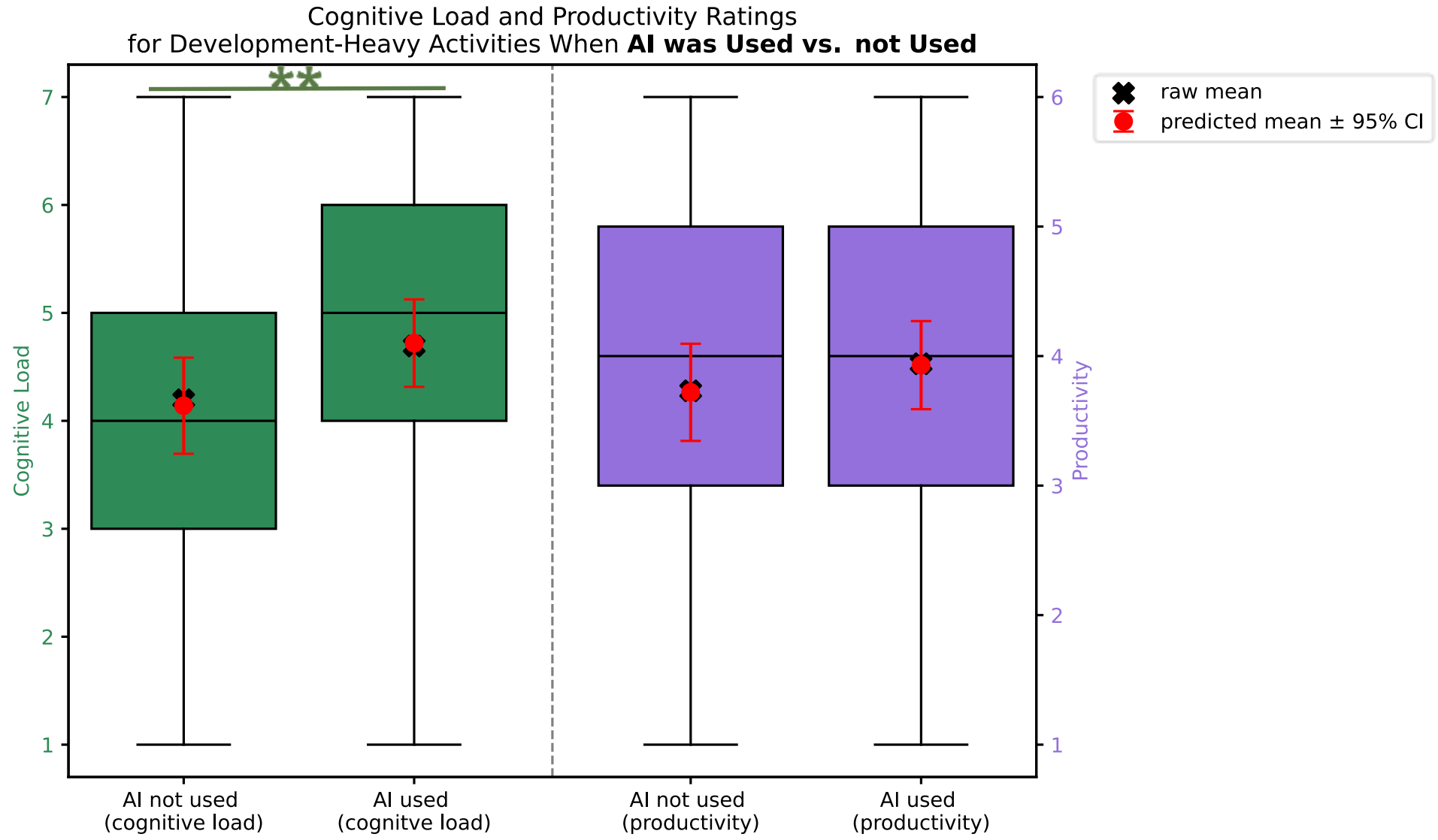

Figure 10: The cognitive load (left and in green) and the productivity ratings (right and in purple) are depicted for the development-heavy activities. The development-heavy activities for which AI was not used is compare with those where AI was used. Asterisks denote a statistical significance (** p≤0.01) difference.

### *5.4.4 AI helpfulness for development-heavy activities*

The follow-up question is whether cognitive load and productivity ratings differ significantly when developers perceive AI as helpful vs not helpful during development-heavy tasks with AI interaction.

(5) cognitive load ~ helpful + (1|participants)

(6) productivity ~ helpful + (1|participants)

with perceived helpfulness of AI (yes vs no) as fixed effect and participants as random effect.

Cognitive load ratings did not differ significantly depending on whether AI was perceived as helpful or not ($EMM_{yes}$=4.70, $EMM_{no}$=4.88, difference=-0.18, SE=0.40, t(98)=-0.44, p=0.659) (Figure 11).

Thus, perceiving AI as helpful during development-heavy tasks was not associated with a significant change in perceived cognitive load.

In contrast, productivity ratings differed significantly with respect to the perceived helpfulness of AI. Tasks in which AI was perceived as helpful were also rated as more productive than tasks in which AI was not perceived as helpful ($EMM_{yes}$=4.00, $EMM_{no}$=3.27, difference=0.73, SE=0.36, t(98)=2.05, p=0.043, d=0.83) (Figure 11

Figure 11: The cognitive load (left and in green) and the productivity ratings (right and in purple) are depicted for the development-heavy activities. The development-heavy activities during which AI was used and found not helpful are compared to those in which AI was used and found helpful. The asterisk denotes a statistical significance (* p≤0.05) difference.). The large effect size indicates that perceiving AI as helpful during development-heavy tasks is associated with substantially higher perceived productivity.

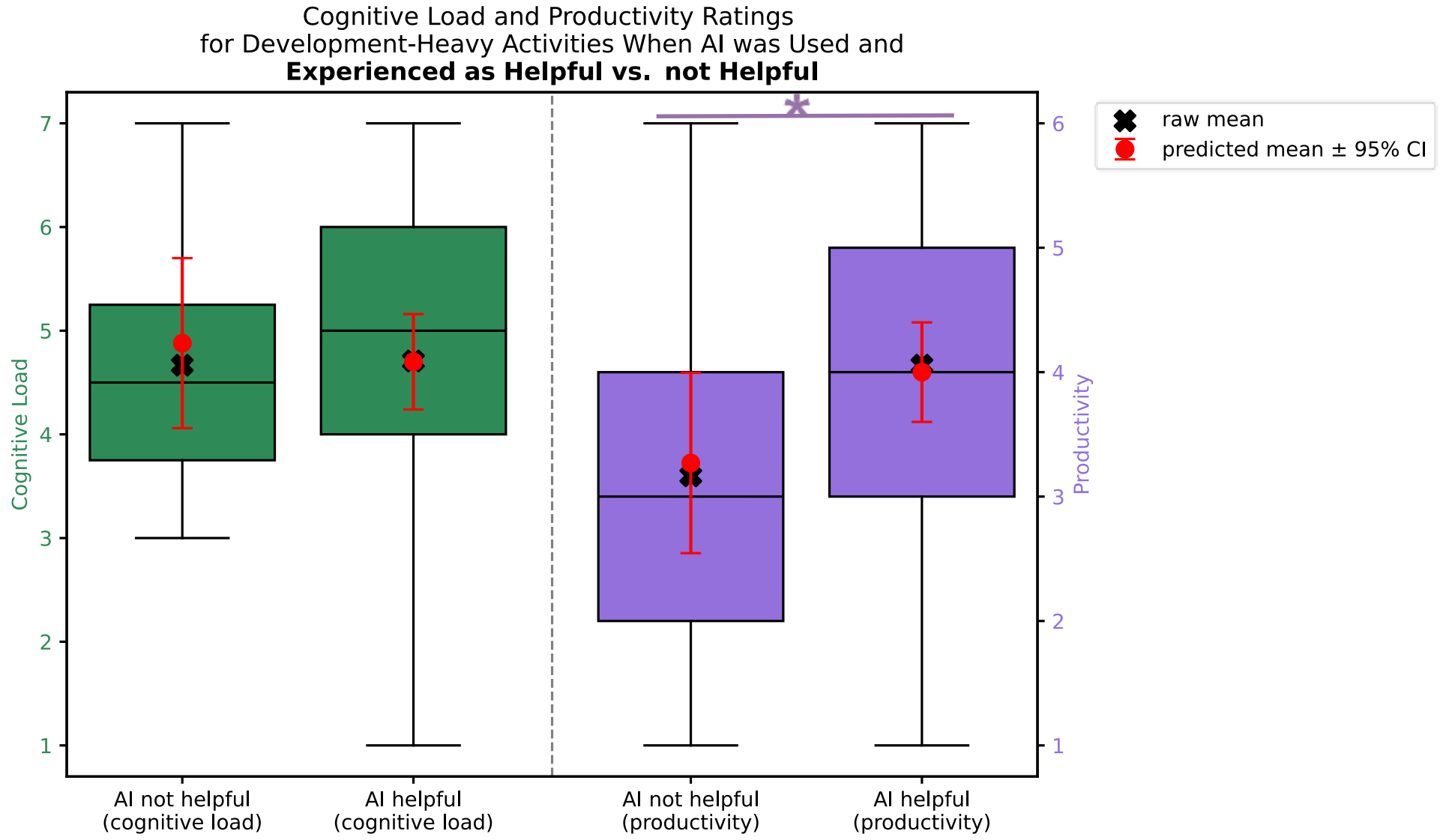

Figure 11: The cognitive load (left and in green) and the productivity ratings (right and in purple) are depicted for the development-heavy activities. The development-heavy activities during which AI was used and found not helpful are compared to those in which AI was used and found helpful. The asterisk denotes a statistical significance (* $p \leq 0.05$) difference.

Among everyday development working tasks, development-heavy activities were the most prevalent and involved the most AI use. Perceived cognitive load and productivity followed the same pattern across task categories, being highest for development-heavy activities, intermediate for collaboration-heavy activities and lowest for other activities. Overall, tasks involving AI were associated with both higher cognitive load and higher productivity than tasks without AI. A closer examination of the development-heavy activities revealed that 1) tasks including AI tools were associated with higher cognitive load but comparable productivity experiences, and 2) tasks in which AI was perceived as helpful showed comparable perceived cognitive load but higher perceived productivity.

### 5.5 Future directions of AI – concerns and opportunities

#### *5.5.1 AI support for software engineering tasks*

All 22 participants selected items from a pre-defined list, indicating which aspects of their job they would most like AI tools to help with. Each selected item was normalized by the total number of items selected per participant. This normalization ensures that participants who selected many items do not disproportionally influence the results, and it gives equal overall weight to each participant. Participants mostly desired AI support for generating tests, followed by root cause analysis, analyzing code for defects, vulnerabilities or optimizations, coding or refactoring code, reducing cognitive load, and writing documentation (Figure 12). None of the participants selected policy management or managing/acquiring permissions. Consistent with participants' perception that AI already assists with monotonous tasks (Figure 2), the most selected items also related to monotonous and time-consuming aspects of their job. Additionally, the desire to reduce cognitive load suggests that participants currently find their work mentally demanding. Participants showed no interest in AI support for security-critical aspects of their job.

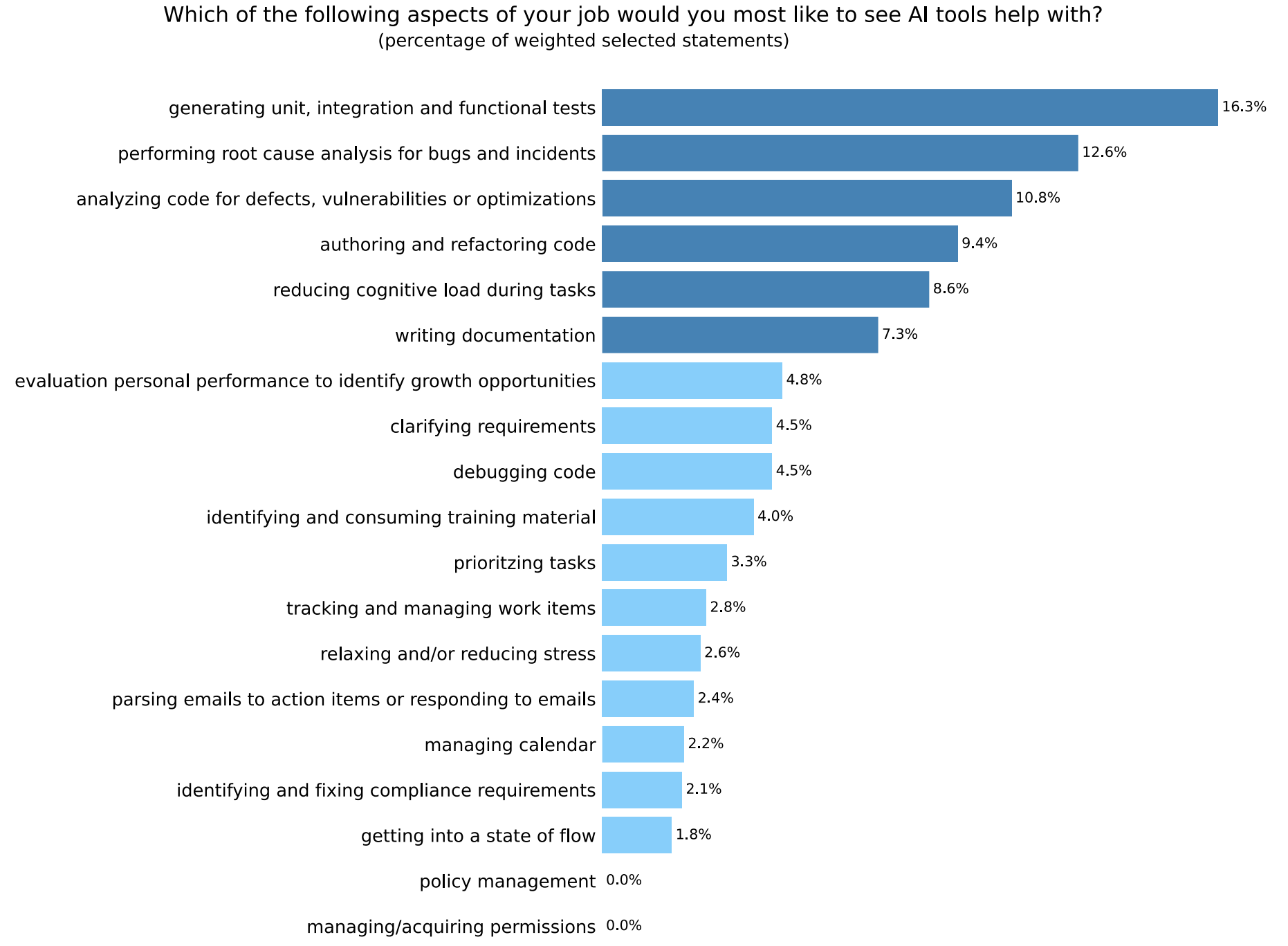

Figure 12: The figure shows which aspects from a given list participants would like AI tools to help with most in their job. Each selection per participant is normalized by the number of aspects selected per participant. The sums of normalized selections per aspect are then divided by the number of participants (22) to receive percentages.

#### *5.5.2 Concerns about AI*

When asked what worries them most about integrating AI into their daily workflow, participants mostly selected introducing defects or vulnerabilities into their work, automating their job away, AI causing their skills to atrophy, and AI being more gimmicky than helpful (Figure 13). None were worried about the time it will take to learn new AI tools or about changing established workflows. Participants were partly concerned about the quality of the output and its impact on their work, including potential skill loss and consequences for their job profile, while still considering AI to be more hyped than useful.

These mixed and somewhat contradictory worries, ranging from AI being too powerful to AI lacking real value, are evident across participants and do not divide them into two factions, as one might assume. However, the absence of concern about learning or workflow changes suggests that the participants are open to innovation.

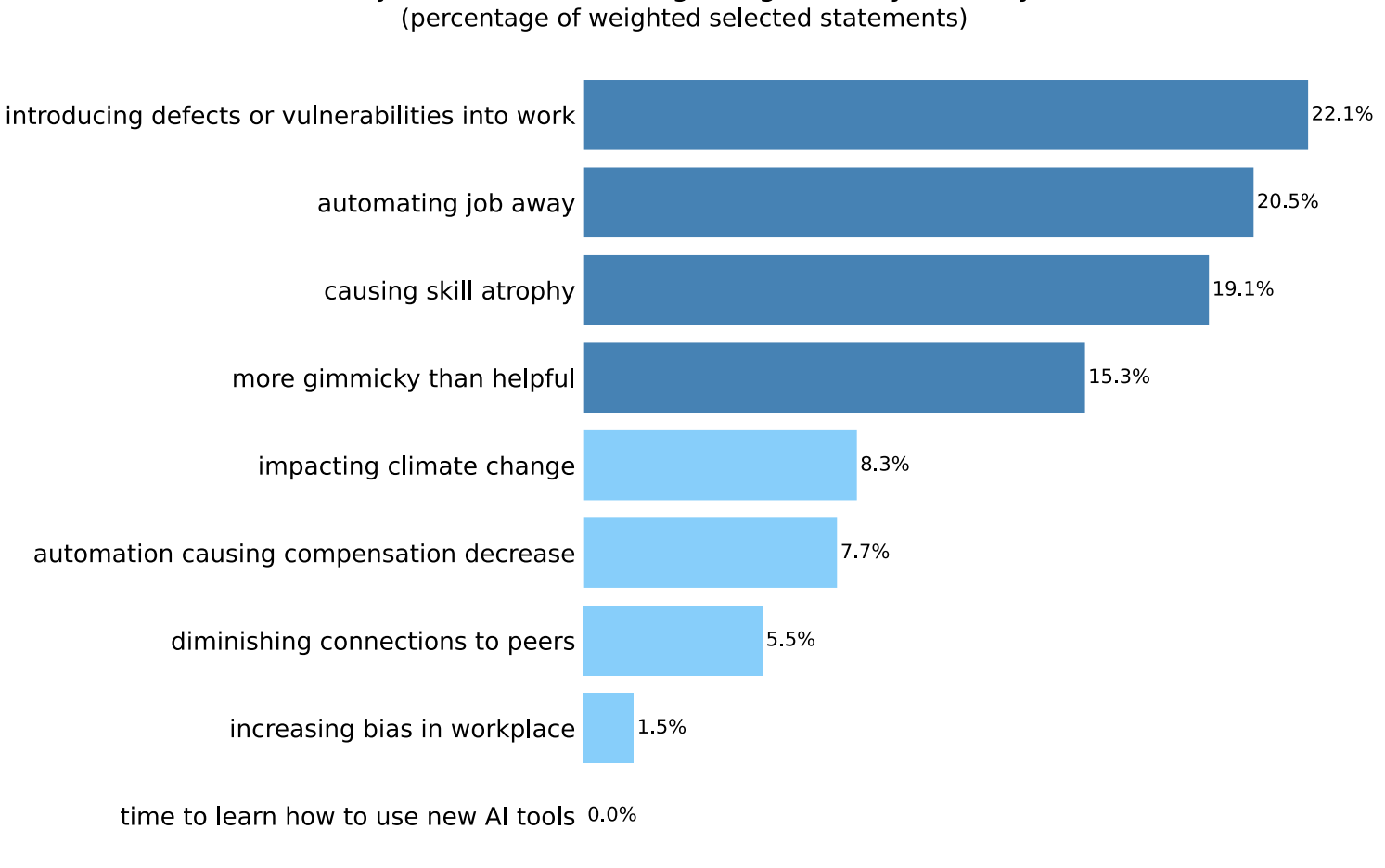

Figure 13: The figure depicts the most pressing worries they selected from a predefined list when integrating AI into their daily workflow. Each selection per participant is normalized by the number of aspects selected per participant. The sums of normalized selections per aspect are then divided by the number of participants (22) to receive percentages.

### *5.5.3 Future with AI*

The participants were encouraged to share additional thoughts on how they see the future of AI in software engineering. From the answers, one could crystallize that the developers see a future where AI takes on especially strenuous and tedious coding-related tasks whilst developers use their reasoning skills and domain-specific knowledge for human- and customer-focused product management (P03, P09, P22, P20, P10, P21). The challenge will be to "define requirements in a non-ambiguous and machine-understandable way" (P02). Some participants express the danger of over-relying on and not completely understanding the AI's output (P13, P14). AI may lead to a loss of a deeper understanding of the code and knowledge in the software engineering domain, according to P14.

These views, like the concerns about AI, give a quite mixed picture. Participants foresee AI taking on parts of their software engineering tasks, but concerns about knowledge loss, overreliance on AI output, and the quality of interaction with AI are also present.

### *5.5.4 Impact of GenAI on team dynamics*

GenAI use can also affect team dynamics. Two questions in the pre-questionnaire aimed to address whether feedback is chosen/preferred from GenAI or team members. According to the median (Md=4), participants agree that they turn more quickly to GenAI tools for feedback than to fellow team members. They, however, disagree (Md=2) with the statement that feedback from GenAI is more helpful for solving a current problem than feedback from team members (Figure 14).

The same two questions were posed in the end-of-workday questionnaire to evaluate the concrete working days during the uncontrolled study period. Here too, participants turned more quickly to GenAI tools for feedback than to their fellow team mates (Md=4), but during the study period, they were undecided (Md=3) whether feedback from GenAI is more helpful for solving a current problem than feedback from team members.

From these results, we can deduce that the quality of feedback provided by team members still somewhat outperforms that of GenAI; however, the hurdle of asking feedback from GenAI is lower than that of asking feedback from team members. As GenAI quality improves, this dynamic may shift further, though diminished connection to peers due to AI does not seem to be the most pressing worry for the participants currently (Figure 13). More extensive studies are needed in this direction in the future.

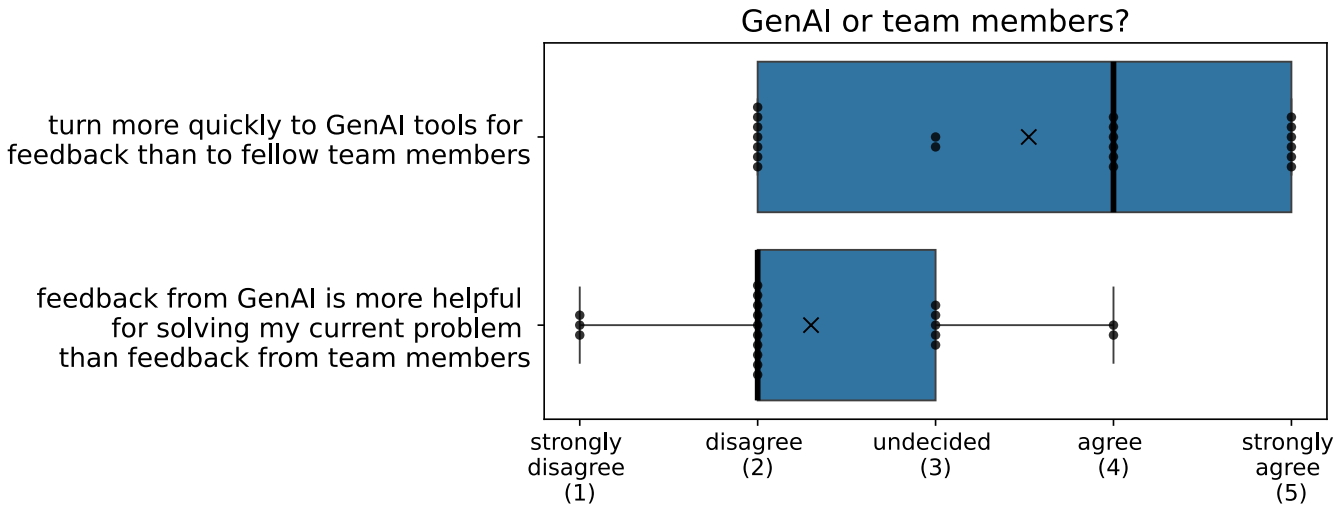

Figure 14: In this figure, two statements were rated by the participants that address preferences for feedback from either GenAI or team members. These are questions from the pre-questionnaire.

#### *5.5.5 Correlations between GenAI interaction and personality*

We were interested in examining whether correlations exist between the five personality traits (extraversion, agreeableness, conscientiousness, emotional stability and intellect/imagination) from the 50-item IPIP test and GenAI interaction. Kendall's tau was used to assess correlations between each trait and participants' self-rated GenAI use at work, duration of use, and GenAI proficiency. Correlations ranged from $|\tau| = 0.03$ to 0.38, indicating weak to moderate associations. Only the correlation between emotional stability and the duration of GenAI use was significantly positive ($\tau$=0.38, p=0.0348), whilst the remaining correlations were not significant. Given the small sample size (n=21), this analysis is only sufficiently powered to detect large effect sizes ($\tau$>0.5) with a statistical power of 1-$\beta$=0.8 and a significance level of $\alpha$=0.05. Non-significant results should therefore be interpreted with caution, since they could reflect limited statistical power rather than the absence of an effect. Emotional stability (with use of GenAI at work: $\tau$=0.26; with GenAI proficiency: $\tau$=0.26) and intellect/imagination (with use of GenAI at work: $\tau$=0.30; with GenAI proficiency: $\tau$=0.35; with duration of GenAI use: $\tau$=0.27) showed weakly positive correlations, whilst agreeableness showed weakly negative correlations (with use of GenAI at work: $\tau$=-0.29; with GenAI proficiency: $\tau$=-0.21; with duration of GenAI use: $\tau$=-0.30). These exploratory results need to be verified with a study of larger sample size.

Software engineers see GenAI as most helpful for tedious tasks that induce cognitive load, like testing, debugging, documentation and analyzing code for vulnerabilities. Support is not wanted for security-related tasks. According to the developers in this study, the work of a developer when using AI will shift towards requiring more high-level reasoning skills and domain knowledge. Key concerns include introducing defects into work, skill loss, and overestimating the usefulness of new tools, although participants remain open to adopting new tools. GenAI's impact on team dynamics and correlations to developers' personalities can be explored in more detail in future studies.

### 5.6 Impact of the study on the view of/interaction with GenAI

#### *5.6.1 Effect of the study on the view of GenAI*

In the post-questionnaire, the participants were asked in an open-ended question whether the (increased) interaction with GenAI during the study period affected how they viewed GenAI. Out of 22, 12 answered in the affirmative, 9 negated, and one participant gave a neutral response. Participants who reported being affected by the study had significantly lower prior GenAI proficiency (U=20, p=0.036, r=-0.4683) and usage (U=24, p=0.0127, r=-0.5061) than those who did not report being affected. They, however, did not differ in work experience (U=52, p=0.5932, r=-0.1125).

From the participants who answered in the affirmative, 5 did not give further explanations on the quality of the affect, but the other 7 participants did:

- "It was very helpful"
- "I increased my use of AI chat, and that is likely to stay."
- "[The study period affected] a bit [how I view GenAI]. I don't use it much to generate code like how GitHub Copilot is able to. I usually use GenAI for conceptual items or debugging."
- "[…] previously I did not have GitHub Copilot installed on my IDE. Now I use it very often […] for code review, test creation, and general questions."
- "[…] I feel I have a better understanding of its capabilities now, and [I will] likely use it more in scenarios where it is productive."
- "I was always using it for my work already, but throughout this study I reflected on it a bit more rather than taking it for granted."
- "[…] it made me realize how much I enjoy no longer writing my own comments and unit tests."

The participants who negated that the study period affected their view on GenAI, also explained that this is because they already regularly use GenAI during their work and know of the benefits and shortcomings or in general find it helpful:

- "[…] Our team uses GenAI often to start and work on problems, so the increased interaction with GenAI stayed mostly consistent with how it was before. It is still helpful to start problems with GenAI, then slowly focus on human opinions and context as the foundation is laid for the solution. It is also helpful to use AI to add finishing touches and refine code."
- "[…] I've been using GenAI tools for the past couple of years and have expanded its use to personal tasks too. My view has been the same, [it is] a very helpful tool."

#### *5.6.2 Change in GenAI interaction*

Upon being asked if their interaction with GenAI will change after this study, 73% (16/22) of the participants affirmed. 23% (5 participants) stated that their interaction will not change, as they are already AI users who frequently or always use GenAI at work, and one participant gave a neutral response.

Some participants want to adapt tool functionalities:

- "I may add my own LLM to the Copilot chat. I have the IDE and chatbot in separate windows and copy and paste a lot"
- "I will turn off inline suggestions after the study"
- "Definitely need the GitHub Copilot license so I can utilize it more at work!"

Others want to change the way they interact by adapting their own behavior:

- "I think I will provide Copilot with more detailed comments for it to generate code I need"

Again others will extend the use cases:

- "maybe I'll use it to generate unit tests and monotonous coding tasks"

- “I will use it more to summarize and brainstorm.”
- “Before I did not realize how GenAI can be used in some tasks. For example, Copilot is already helpful with coding, documentation, and testing. I felt skeptical and dismissive towards some tools […]. I also may start looking for GenAI productivity tools. Due to the increased awareness of cognitive load during the study, I noticed that a great deal of cognitive load can come from simply management of several windows, tools, and tasks at once.”

> Over half of the participants reported that the study influenced their view of GenAI, especially those with less prior experience, leading to plans for broader or more refined use. Frequent users saw little change, as they were already aware of GenAI’s strengths and limitations. 73% of all participants said they would adjust their use of GenAI going forward, from tool setup to applying it to new tasks. Overall, the study encouraged greater awareness and more intentional use of GenAI in software development.

## 6 DISCUSSION

**RQ1: Overall satisfaction with GenAI, particularly for efficiency gains and reduced effort during monotonous tasks, with dissatisfaction arising from the misalignment of model capabilities, appropriate tasks and users’ GenAI interaction expertise.** Overall, the results of the pre-questionnaire indicate that developers primarily associate AI with efficiency gains and workload reduction rather than with intrinsic work value, such as work fulfilment. Participants largely agreed on the usefulness of AI for monotonous tasks, productivity improvements, and as a helpful pair-programming tool, indicating they value GenAI as an assistive technology that reduces the effort of their work. In contrast, responses regarding creative tasks and work quality were more neutral or mixed, suggesting uncertainty about AI’s contribution beyond efficiency gains. Nearly half of the participants disagreed that AI enhances work fulfilment and ratings are on average neutral regarding AI improving motivation at work. This indicates that AI support for productivity gains does not necessarily translate into use with intrinsic motivation. A possible explanation is that GenAI tools are partly automating aspects of software development which developers find enjoyable, such as hands-on coding. Participants observe an increasing shift towards higher-level problem definition and solving, and towards critical evaluation of AI-generated outputs, which may contribute to a feeling of less fulfilment. While these findings align with prior studies that report perceived productivity and efficiency gains during AI use [1, 9, 29, 45], evidence for improvement of intrinsic work value is less consistent. While Ngwenyama et al. [29] found that their participants experienced an increased sense of flourishing at work, the findings in the present study do not indicate this. Overall, GenAI is perceived more as a means of enhancing efficiency than as a contributor to intrinsic work fulfilment.

The participants' perceptions of AI usefulness for monotonous tasks and improved productivity during work align with their actual usage patterns of GenAI tools. GenAI was most frequently used for writing code, writing and running tests and debugging, which are coding-related tasks that consume most time during the developers’ regular work week. This distribution of working tasks over a week is largely consistent with our findings from a study conducted one year earlier with a similar cohort of professional developers [6], where coding, meetings, and debugging also accounted for most of the working time. A notable difference in the present study is that testing consumes more of developers’ working time. This shift may reflect differences between participant cohorts or indicate an increased need for testing and verification as more and more code is AI-generated. The latter explanation is supported by findings of our present and prior research, which shows that developers need to put effort into reviewing and validating AI outputs [27]. Overall, developers seem to integrate AI into their work where it offers the greatest time and effort savings, and where the assistance provided by AI can be seamlessly embedded into their workflows. On the other hand, GenAI tools are used less frequently for creative or text-oriented tasks, such as brainstorming or summarizing, which suggests that these application scenarios are perceived as less useful by the developers.

The results on developers' satisfaction with GenAI further support the interpretation that developers primarily engage with GenAI in pursuit of efficiency benefits. Participants reported high overall satisfaction with their interactions with GenAI, with most indicating that they were satisfied or very satisfied, and none expressing outright dissatisfaction. Open-ended responses offer valuable insights into the drivers of satisfaction and sources of dissatisfaction. Satisfaction was largely associated with output quality and work efficiency, particularly for tasks that are repetitive, time-consuming, or structured, such as writing unit tests, debugging, and writing boilerplate code. Work efficiency-related statements were consistently mentioned as a positive aspect of GenAI, underlining the aim of GenAI use to improve development efficiency. These impressions align with prior research: Copilot was found useful for tasks such as testing and boilerplate code generation [1, 9], and work efficiency gains and time optimization were reported in field and controlled studies [29, 32, 45].

Interaction quality also contributed to satisfaction, as participants valued GenAI for providing new ideas, alternative solutions or initial suggestions that serve as starting points for problem-solving, especially when prior knowledge was limited. These findings align with those of Vaithilingam et al. [38]. Sources of dissatisfaction were mostly related to model limitations, including inaccuracies, hallucinations, prompt sensitivity and the need for human oversight, manual adjustments and verification. In the work context, internal models are sometimes outdated and/or insufficiently adapted to business domain knowledge. This limitation, however, can be partly attributed to the fast-paced developments of more accurate models and to new tools having to be integrated into the organizational workflows, so a certain latency is unavoidable. These challenges echo prior reports of reliability issues and the cognitive effort required to validate AI output [9, 34, 38], indicating that the effective use of GenAI still requires substantial user effort and expertise. Overall, these statements on satisfaction and dissatisfaction with GenAI indicate that developers' satisfaction is closely tied to the alignment between the capabilities of underlying models, working tasks at hand and the users' experience with GenAI interaction. GenAI is primarily used to improve work efficiency for monotonous tasks, particularly when such improvements can be achieved with minimal interaction overhead and seamlessly with little human intervention.

**RQ2: Preferred interaction type depends on use cases.** During the controlled sessions of this study, in-code suggestions were used more frequently and were associated with higher ratings of satisfaction and perceived productivity compared to chat-based interaction. The generation of code at the cursor position within the code enables a rapid, low-effort code generation within the existing workflow, which can explain this preference. As could also be seen in the screen recordings, in-code suggestions are used more on a trial-and-error basis, with a lot of code being generated but also discarded again. This exploratory interaction style, together with the amount of generated code, may induce a sense of achievement, thereby increasing perceived productivity and satisfaction. The limited code context for generating in-code suggestions (relying on the immediate surrounding code) appears well-suited for less complex tasks that require small changes to the code.

Due to the continuous dialogue-like nature of the chat interaction, which also takes a larger codebase into account for its output, chat is preferred and applicable for more complex tasks like debugging and non-coding tasks, such as brainstorming and writing summaries. In these situations, developers value explanations for the output and the continuous interaction. Together, these findings suggest that developers' interaction preferences with Copilot are task-dependent: in-code suggestions support rapid code-level exploration, while chat interactions support higher-level reasoning and problem understanding.

**RQ3: Accuracy, efficiency and perceived workload were positively affected by using only one interaction type during a task compared to no Copilot interaction or both interaction types.** The above-discussed findings revealed a lack of clarity among developers about how to effectively interact with GenAI systems, which can cause dissatisfaction. Not knowing the amount and type of background knowledge required by the model, or choosing a non-optimal interaction type, can lead to suboptimal results.

Our results showed that using one interaction type during a task, i.e. either in-code suggestions or chat, increases efficiency (measured by task duration) and reduces perceived workload (captured via the raw NASA-TLX score) compared to no Copilot use during the controlled sessions of this study. Task completion, i.e. accuracy, benefited most from chat use. For interaction with chat, more high-level thought is required to formulate the prompt, but the perceived workload is still lower than not using GenAI at all. If a prompt is formulated in a comprehensive way for the model to generate the desired output, fewer interactions are needed to receive a result [7].

However, combining two interaction types does not reduce task duration or perceived workload compared to not using GenAI. This result can indicate that switching between interaction types may result from one type being suboptimal for solving the problem at hand or from evolving task demands. This switching and the additional consideration of how to best solve the problem given the available GenAI interaction options can lead to increased cognitive overhead, resulting in efficiency and workload levels comparable to those observed without GenAI support.

These findings align with and extend prior research on GenAI-assisted software development by clarifying how interaction choices shape observed productivity and experience effects. While earlier studies reported improvements in productivity or efficiency when using tools such as GitHub Copilot in general [1, 32], our results paint a more nuanced image, namely that these benefits depend on the interaction type chosen for a specific task rather than on Copilot usage per se. Moreover, the accuracy benefits of chat-based interaction complement earlier qualitative findings that developers particularly value GenAI for higher-level reasoning and explanation tasks [9, 34, 38].

➡ **Rules of thumb for interaction type choice.** Taken together, our results suggest that prior thought should be invested into choosing the appropriate interaction type for a specific problem at hand, rather than switching between interaction modes during a task. Once a GenAI tool such as Copilot has been selected, the following guiding questions may support this decision:

- Is the task primarily coding or non-coding related?
- How much context knowledge of the codebase does the model require to effectively assist?
- Are explanations of the output required?

As a rule of thumb, coding-related tasks that require little code context and do not require extensive explanations beyond inline comments tend to be well-suited for in-code suggestions. Tasks can include boilerplate coding, documentation and testing. For non-coding-related tasks that cannot be supported by in-code suggestions or for tasks that require larger contextual knowledge of the codebase and where explanations are desired, chat-based interaction should be preferred. Tasks may include brainstorming, writing summaries, debugging, retrieving information from a codebase, and generally, more high-level, creative, or complex tasks. Selecting the appropriate interaction type for a task upfront may help avoid frequent interaction type switching and thereby reduce minor workflow interruptions, which can improve efficiency and lower perceived workload.

**RQ4: AI interaction during everyday development-heavy activities involves a higher perceived cognitive load and comparable perceived productivity than non-AI use.** The evaluations of the developers' everyday working tasks during the uncontrolled period of the study indicate that both tasks and AI use significantly influence developers' perceived cognitive load and productivity. Concerning the tasks, development-heavy activities were perceived as most cognitively demanding and were simultaneously associated with the highest perceived productivity. Collaboration-heavy activities and other activities showed lower cognitive load and productivity ratings, but also a consistent coupling between cognitive load and productivity. These results align with the flow theory, which suggests that cognitively demanding tasks can be those in which people perceive themselves as accomplishing more [10]. This helps to explain the connection between perceived cognitive load and productivity.

AI use was most prevalent among development-heavy activities compared to collaboration-heavy or other activities. However, when AI was used during non-development-heavy activities, the interaction was experienced almost always as helpful. In line with the results of the controlled sessions in this study, one could argue that the AI use case was more thoroughly planned and considered than the most common use case in programming. Therefore, a more conscious use of the AI tool at hand led to it being perceived as helpful. A more conscientious use and exploration of AI tools was reported by some participants in open-ended questions after the study, which supports this interpretation further.

The higher amount of AI interaction among all participants and the mixed perception of helpfulness among eight participants during development-heavy tasks motivated a more detailed examination of cognitive load and productivity during these tasks. Tasks with AI interaction showed higher perceived cognitive load and comparable perceived productivity to tasks without AI use. Several explanations are plausible: one possibility is that AI support was chosen for more complex tasks, which inherently lead to higher cognitive load due to the nature of the tasks and independently of the AI interaction. Perceived productivity remained comparable because AI support, even if imperfect, helped during the complex tasks. However, participants' ratings and reports of using AI during monotonous, routine, time-saving tasks make this explanation less likely. Assuming, therefore, comparable task difficulty, the increased cognitive load may reflect the additional overhead of interacting with AI in terms of switching between AI tools and interaction types within the tools, or needing to verify and adjust AI output. Thus, even when AI is applied to routine tasks, the interaction and verification overhead can outweigh reductions in task-related cognitive demand. Perceived productivity remains stable because developers are still able to complete the tasks at a similar level as without AI, despite the extra mental effort. This interpretation is further supported by the observed results that AI interaction, which was perceived as helpful, has comparable cognitive load ratings whilst increased productivity ratings compared to AI interactions rated as not helpful. The effort of interacting with AI is similar, but perceived productivity is higher when the output is perceived as helpful.

In sum, AI interaction tends to increase cognitive load, but when the AI output is perceived as helpful, it can also lead to higher perceived productivity without additional cognitive load. In development-heavy tasks, cognitive load appears to arise primarily from interacting with AI, whereas perceived productivity gains depend on the output quality. The mixed ratings of the participants when asked if it is less mental effort when using AI at work than without also indicate that AI use is not per se a cognitive relief for all developers.

**RQ5: Openness to learning new AI tools, accompanied by worries and expected changes in software developers' work.** The current AI landscape is characterized by a mixture of enthusiasm for and satisfaction with AI, as well as by a realization of its limitations and associated challenges. Software developers in this study show openness to learning new AI tools and adapting their workflows accordingly, but they also see the risks of introducing defects into their work, losing acquired skills, and needing fewer developers in the future due to AI. While developers wish that AI would primarily support them by taking over especially strenuous and tedious tasks, they envision their future roles focusing on high-level problem definition in an AI-understandable form, reasoning skills, domain-specific knowledge and on human and product management. While direct empirical evidence showing that developers lose acquired skills specifically due to AI use is not yet established in the software engineering literature, leveraging AI for efficiency and productivity gains may reduce the number of developers who maintain deep expertise and full oversight of complex codebases. Ensuring adequate verification and feedback to AI tools in case of errors may increasingly depend on a smaller subset of highly expertized developers in the future, highlighting the importance of preserving domain knowledge and critical evaluation skills.

This study also briefly explores the impact of GenAI on team dynamics. With the introduction of AI tools in the workplace and with further improvements of AI tools, team dynamics may change. The results indicated that developers turn more quickly to GenAI tools than to fellow team members, but they currently tend to find their colleagues' feedback more helpful to solve their problem. Strengthened by Stray et al.'s findings [38], our results can be interpreted as showing that developers turn to GenAI for less complex feedback so as not to disturb their colleagues, while they turn to colleagues when requiring trustworthy feedback on more complex problems that require domain knowledge. The results suggest that AI complements rather than replaces team interactions, although future improvements in AI quality could alter this balance.

Exploratory analyses of personality traits showed only weak to moderate associations with GenAI use: emotional stability and intellect/imagination were positively related to self-reported AI use and proficiency, whereas agreeableness showed a weak negative correlation. Given the small sample size, these results should be interpreted cautiously and may reflect limited statistical power rather than definitive patterns. While prior work has not extensively examined personality in the context of GenAI-assisted development, research suggests that individual differences can influence how tools are used and perceived [2, 27, 38, 40, 41]. Although preliminary, our findings suggest that personality could help explain variation in AI adoption, interaction strategies, and perceived usefulness. Future studies with larger sample sizes could more extensively investigate the impact of developers' personality on their interactions with AI tools.

Taken together, our findings show that while developers are generally open to learning about tools and integrating them into their work, they observe both perceived benefits and potential risks. AI appears to complement existing workflows and team structures rather than replace human judgment, with individual differences in personality potentially influencing how developers engage with and benefit from these tools. Future research should continue to explore these dynamics, particularly as AI capabilities evolve and become more integrated into software engineering practice.

**RQ6: Positive impact of conducting studies in a firm setting for the individual and for the firm.** The subjective open-ended answers of the participants indicate that conducting this study had a positive impact on their views of and interactions with GenAI. While just over half of the participants reported a change in how they view GenAI, 73% of the developers expressed intentions to adjust their interaction with GenAI going forward. Especially developers with lower GenAI proficiency and usage were significantly more likely to report a change, whereas users mostly reported stable views, frequently explaining this by their already existing knowledge of GenAI's benefits and shortcomings.

These findings suggest that the observed impact of GenAI on developers' interactions does not stem solely from increased exposure to GenAI, but also from structured, more reflective interaction with AI technology during simulated work tasks over the study period. Participants' responses indicate reflection on their AI use during the study, with some planning to adapt or expand their AI use or refine their interaction strategies, such as adjusting tool configurations, disabling certain features, or providing more detailed contextual input. For the individual developer, the study indirectly encouraged more reflection on intentional and differentiated use of GenAI in software development workflows.

At the organizational level, these findings highlight the value of conducting empirical studies with the active involvement of employees, surpassing mere questionnaire-based evaluations. Similar study setups could be used in firm settings not only to actively train employees on when and how to use new technology and how to assess the validity of its outputs, but also to provide firms with insights into how such tools could be actually used in practice and in which situations they are most beneficial. This aligns with McRae et al. [20], who emphasize the need for quality control and sound employee judgment when working with GenAI. In this context, workshop- or hackathon-style formats embedded in real work settings can be an effective way for firms to introduce tools such as GitHub Copilot. Such formats allow developers to experiment, reflect, and exchange experiences, while simultaneously enabling organizations to evaluate the practical benefits, limitations, and interaction patterns associated with GenAI tools systematically.

## 7 CONSIDERATIONS ON VALIDITY

### 7.1 Construct validity

The part of the study described in this paper aimed to capture developers' interaction with GenAI in a natural work environment as realistically as possible. The study was conducted directly at SAP and included controlled and uncontrolled phases. Whilst the controlled sessions can arguably not be considered a day-to-day work scenario, the work surrounding was familiar to the developer, making the setup more realistic than a laboratory setting.

Analyzing subjective data gathered via questionnaires and objective data from screen recordings together aimed to give more nuanced and holistic insights into the developers' experience with GenAI interaction. Questionnaires were mainly based or slightly adapted from existing questionnaires to ensure comparability and quality of the posed questions.

The coding-related tasks during the controlled sessions were chosen from the HumanEval-X benchmark dataset, thereby providing standardized tasks. Everyday working tasks were documented by the participants during the uncontrolled study period, thereby adding natural tasks to the study setup.

### 7.2 Internal validity

The realistic work environment naturally introduced confounding variables. To address this, we incorporated both controlled and uncontrolled phases in the study design and gathered contextual data via questions about participants' mood, work environment, and unusual events. This additional information helps contextualize the results and documents impeding factors, but further confounds may still be impacting the interaction.

### 7.3 External validity

The general ecological validity of the study is high since professional software developers participated in the study in their accustomed work environment. The results do not need to be transferred from a laboratory to a natural work context.

However, questions may arise about the generalizability of the study's findings from the controlled sessions. During the sessions, tasks were solved only with a single GenAI tool - GitHub Copilot - and the developer-GenAI interactions were only analyzed within this context. While interaction behavior and user experience can vary between tools, popular GenAI coding assistants also offer a combination of in-code suggestions and natural language (chat-based) interface, such as Cursor or Amazon Q Developer. We therefore expect that our results are transferable to other coding assistants, though further research is needed to validate this assumption.

During the uncontrolled period, participants were free to choose any AI tool for interaction. Comparing results from this uncontrolled period with the controlled sessions and interpreting them together supports the generalizability of our findings.

All controlled coding-related tasks were completed in Java, a language participants were familiar with. Future studies need to investigate whether the GenAI interaction experience and behavior varies between programming languages. Small variations may be expected since tool performance can vary with programming language [28] and acceptance rates whilst using Copilot can also vary between programming languages [1].

We acknowledge that perceived cognitive load, perceived productivity, and, in general, perceptions are highly individual, experience- and task-dependent. A nuanced yet general trend across developers is portrayed in this paper, and the subjectivity is strengthened by objective behavioral results. In future work, we will also evaluate the physiological and keyboard and mouse movement data collected in the study as additional personal yet objective data sources.

## 8 FUTURE WORK

Future research will further explore developer-GenAI interactions during the controlled and uncontrolled study phases of this study, with a focus on cognitive load measured through physiological signals, keyboard activity, and mouse usage. Additional studies could investigate how GenAI affects team dynamics and how these effects relate to individual personality traits. The current study design could also be adapted to examine interactions with different or multiple code generation tools and large language models, as well as the influence of different chat interaction modes (e.g., ask, edit, or agent).

Our study results have shown that human verification of the model output is still vital, and response quality is not always evident. Currently, there is limited feedback available on the correctness of outputs or the confidence level of the model output. Current and future research needs to address this problem in the field of explainable and trustworthy AI to improve user trust and transparency of GenAI systems.

Building on prior work by Barke et al. [2], which distinguishes between acceleration and exploration modes in Copilot use, future studies could investigate how GenAI tools might adapt their suggestions to the developer's current mode to reduce the risk of interrupting the programming flow and reduce cognitive load.

In this study, we considered the perceived workload induced by different interaction types (in-code suggestions vs chat prompting) across task types. Future work aims to leverage physiological data from wearables to provide continuous, real-time cognitive load feedback within an IDE. This could enable adaptive adjustments of interaction types and modes by both the developer and the GenAI tool in the future. Following Mozannar et al. [27], we also plan to explore the integration of human-in-the-loop feedback mechanisms to better tailor GenAI interactions to developer needs.

Next to GenAI assistants, Agentic AI systems are being increasingly integrated into software development workflows. Although this study focused on developers' experience during GenAI interaction, several findings may be transferable to the developer-agent interaction. Instead of switching between GenAI interaction types, minor interruptions and increased workload could result from switching between agents. Similarly, the need for systematic and intentional interaction strategies will also apply to effectively orchestrate agents. Furthermore, the developers' anticipation of higher-level planning, reasoning, and task oversight for the future with AI aligns with the vision of agent-supported software development. In such settings, cognitive resources will likely be directed to understanding agent decisions, validating them, and adapting the agents' functioning. Next to examining developers' interaction with AI, future research may focus on developers' oversight of AI systems. Empirical studies are needed to determine whether the interaction patterns and cognitive demands observed in the context of GenAI interactions are transferable to interactions with agentic AI systems.

## 9 CONCLUSION

In this paper, the combination of subjective and objective data sources and of controlled and uncontrolled study periods enabled a realistic, holistic, human-centered and detailed analysis of professional developers' interaction behavior and experience with AI tools. We found that developers are generally satisfied with AI during their work and that productivity gains are a strong motivator for using AI tools. However, intrinsic motivation to use AI tools that would lead to higher work fulfilment, or to experience the interaction itself as a motivation at work, remains limited. A more detailed look at the developers' interactions with Copilot during the controlled sessions and self-reports in this study suggests a mainly explorative and intuitive interaction with AI. The benefits of Copilot use were found when a single interaction type was used for a task, whereas switching interaction types led to results comparable to not using Copilot. Therefore, we propose a rule of thumb to motivate a more systematic and planned approach and the use of AI tools depending on the nature of tasks at hand. Developer-AI interaction optimization that leads to 1) less task and interaction type switching, and 2) increased output quality and understandability of model output could increase efficiency benefits and reduce developers' cognitive load. Developers foresee a change in the development work with the increasing integration of AI into workflows such as tasks requiring more high-level planning, reasoning and context knowledge. Developers in this study expressed openness to learning new AI tools but also mentioned concerns such as skill atrophy. As AI quality and integration into workflows increase, team dynamics may be affected with developers turning faster to AI for feedback and relying more on AI than colleagues' judgement, highlighting the need to observe this potential shift carefully. As GenAI usage varies widely among developers in terms of frequency, task coverage, and interaction style, the reflective engagement encouraged by the study itself highlights the value of structured yet flexible study formats. They can help both individuals and firms move towards a more informed, intentional, and productive use of and interaction with GenAI in software development.

**ACKNOWLEDGMENTS**

We would like to sincerely thank all SAP developers who participated in this study for their time, effort, valuable insights and interest in this research. The work of Charlotte Brandebusemeyer is funded by the HPI-SAP Research Program.